\documentclass{jfm}
\usepackage{graphicx}
\usepackage{grffile} 
\usepackage{epstopdf, epsfig}
\usepackage{caption, subcaption} 
\usepackage{amsmath}

\usepackage{xcolor}
\usepackage{latexsym}
\usepackage{hyperref}

\usepackage[demo,abs]{overpic}
\usepackage[export]{adjustbox}

\newtheorem{theorem}{Theorem}

\shorttitle{Local Koopman spectrum and properties}
\shortauthor{W. Zhang, M. Wei}

\title{A theoretical framework for Koopman analyses of fluid flows, part 1: local Koopman spectrum and properties}

\author{Wei Zhang\aff{1},
 Mingjun Wei\aff{1}
 \corresp{\email{mjwei@ksu.edu}}
}

\affiliation{\aff{1}Mechanical and Nuclear Engineering, Kansas State University,
1701B Platt St, Manhattan, KS 66506, USA}

\begin{document}

\maketitle

\begin{abstract}

Local Koopman spectral problem is studied to resolve all dynamics for a nonlinear system. The proposed spectral problem is compatible with the linear spectral theory for various linear systems, and several properties of local Koopman spectrums are discovered. Firstly, proliferation rule is discovered for nonlinear observables and it applies to nonlinear systems recursively. Secondly, the hierarchy structure of Koopman eigenspace of nonlinear dynamics is revealed since dynamics can be decomposed into the base and perturbation parts, where the former can be analyzed analytically or numerically and the latter is further divided into linear and nonlinear parts. The linear part can be analyzed by the linear spectrums theory. They are then recursively proliferated to infinite numbers for the nonlinear part. Thirdly, local Koopman spectrums and eigenfunctions change continuously and analytically in the whole manifold under suitable conditions, derived from operator perturbation theory. Two cases of fluid dynamics are numerically studied. One is the two-dimensional flow past cylinder at the Hopf-bifurcation near the critical Reynolds number. Two asymptotic stages, flow systems around an unstable fixed point and a stable limit cycle were studied separately by the DMD algorithm. The triad-chain and the lattice distribution of Koopman spectrums confirmed the proliferation rule and hierarchy structure of Koopman eigenspace. Another example is the three-dimensional secondary instability of flow past a fixed cylinder, where the Fourier modes, Floquet modes, and high-order Koopman modes characterizing the main structure of the flow are discovered.

\end{abstract}

\begin{keywords}
nonlinear dynamic system, Koopman operator, local Koopman spectrum
\end{keywords}

\section{Introduction}

Nonlinear dynamic systems play an important role in scientific research and engineering applications because most systems are inherently nonlinear. Unfortunately, nonlinearity makes a system much more complicated. On one hand, a nonlinear dynamic system can admit various complex solutions such as stable or unstable nodes/spirals/limit cycles or even multiple of them. Sometime there can be strange attractor, or even chaos solution~\citep{strogatz2018nonlinear}. On the other hand, nonlinearity can create more complicated modes by self-interacting of flow structures. The modulation of modes as well as the dynamics are all attributed to nonlinearity, for example, the facts that base flow and the growth rate change in a Hopf-bifurcation process were attributed to nonlinearity~\citep{landau1959course}. Different from the well-studied linear dynamics, a nonlinear process remains a challenge in many situations.

Recently, there are emerging efforts to analyze the dynamics by a spatial-temporal decomposition in a data-driven manner. Researchers apply various decomposition techniques on the outcome of the systems to analyze their dynamic properties, especially in the fluid regime. These techniques are efficient in reconstructing dynamics in low-dimensional modeling. For instance, proper orthogonal decomposition (POD) was found to capture coherent structures of turbulent flow~\citep{holmes1996turbulence} and rebuild dynamics efficiently~\citep{lucia2004reduced}. Dynamic mode decomposition (DMD) based on eigendecomposition is also efficient to capture dynamics for many fluid problems~\citep{schmid2010dynamic}. The above decomposition techniques provide exact solutions for linear systems, while approximates well for certain nonlinear systems. Other than the mentioned data-driven manner using a set of dynamics-induced spatial bases, other fixed spatial bases derived from Sturm-Liouville theory~\citep{marchenko1977sturm}, such as Fourier series, Chebyshev series, Bessel series or other trigonometry or polynomial series, are found efficient in many applications~\citep{kim1987turbulence,orszag_1971,andrews1992special,mathelin2005stochastic,jaganathan2008realistic}.

Besides the above decomposition techniques, the spectral decomposition of the linear, infinite-dimensional Koopman operator defined on a nonlinear dynamic system provides a promising decomposition for the nonlinear system~\citep{mezic2005spectral,rowley2009spectral}. Koopman operator has a long history for nonlinear dynamics analysis.~\citet{koopman1931hamiltonian} first introduced a linear operator, which is known today as Koopman operator, to `transform' a nonlinear system to be evoluted by a linear operator.~\citet{neumann1932proof} found the fixed point (if it exists) of the Koopman operator defined on gas molecules dynamic system was an efficient device to determine if the system is ergodic. For instance, the equilibrium of thermodynamics (the zeroth law of thermodynamics) can be defined at the molecular level by finding such a fixed point of molecular dynamics despite the constantly moving gas molecules~\citep{neumann1932physical}. Later, the spectral theory was developed for operators, since linear mapping on the infinite-dimensional linear space naturally raises spectral problem~\citep{reed1972methods,dunford1963linear,dowson1978spectral}. Therefore, the recent works of spectral Koopman decomposition have received a lot of attention.

However, spectral theory of infinite-dimensional operator is more complicated than the finite-dimensional counterpart, the matrix spectral theory. For example, it is known bounded operators have bounded spectrums, while spectrums for unbounded operators remain undetermined~\citep{reed1972methods}. Therefore, current applications~\citep{mezic2005spectral} focused on periodic or ergodic systems, which define the unitary operator, one type of bounded operators. Unfortunately, extending operator spectral problem to a general nonlinear system is difficult from current authors and a few others' experiences. For instance, inconsistency was discovered when approximating Koopman spectral decomposition using numerical DMD technique~\citep{page2019koopman}. Or, to address the spectral problem of operator, artificial families of discontinuous unimodular Koopman operator were constructed for a nonlinear system~\citep{mohr2014spectral}.

To address the mentioned difficulties, a local Koopman spectral problem on nonlinear dynamic systems is defined and studied in this work. This is the first step current authors have taken to understand the complicated behavior of nonlinear dynamics. They are the key to understand the properties of nonlinear dynamics, which will be explored extensively in part 2 of this paper, where a theoretical framework for both linear and nonlinear dynamics will be developed.

In part 1, Koopman spectrums of various linear systems are introduced in \S 2. Starting from a linear time-variant (LTI) system to a general linear time-variant system, Koopman spectrums are found to change from constants to time-dependent values. Then state-dependent Koopman spectrums for nonlinear dynamic systems are introduced in \S 3, generalizing the local spectrum discovered in LTV systems to nonlinear autonomous systems. Several important properties of Koopman spectrums of nonlinear dynamics are introduced. They are the proliferation rule, the hierarchy structure, and the continuity property. In \S 4, Koopman modes of the full-state variable are discussed for different linear systems, while modes for nonlinear systems will be discussed in part 2. DMD, the data-driven technique to realize Koopman decomposition, is examined in \S 5 for its applicability to various dynamic systems, such as LTI, LTV, or asymptotic nonlinear systems. Finally, two numerical examples are presented in \S 6. The primary and secondary instabilities of flow passing a fixed cylinder are studied to support the discovery of local Koopman spectrums.

\section{Koopman spectral problem for linear dynamic systems}


Considering a discrete dynamic system evolving on a manifold $\mathcal{M}$ such that, 
\begin{equation}\label{eqn:discretesystem}
\boldsymbol{x}_{k+1} = f(\boldsymbol{x}_k),
\end{equation}
$\boldsymbol{x}_k \in \mathcal{M}$. $f$ is a map from $\mathcal{M}$ to itself, which evolves the dynamics one step forward, and $k$ is an integer time step. 

An \textit{observable} $g$ is a function defined on the manifold $\mathcal{M}$, $g: \mathcal{M} \rightarrow \mathbb{R}$. The use of observable to study the dynamics introduces flexibility for many applications, but more importantly, to transfer dynamics to its functional space whose importance will be made clear in part 2. Besides, the trajectory of $\boldsymbol{x}(t)$ may not always be available or of interest for some applications. For example,~\citet{lasota2013chaos} studied the probability density function (PDF) instead of the sensitive and chaotic trajectory to understand the dynamics of nonlinear deterministic systems.

The \textit{Koopman operator} $U$ is an infinite-dimensional linear map defined on the observable. It evolves the observable one-step forward by
\begin{equation} \label{eqn:Koopmandef}
Ug(\boldsymbol{x}_i) = g(\boldsymbol{x}_{i+1}) = g(f(\boldsymbol{x}_i)).
\end{equation}
The Koopman operator is a linear operator, since
\begin{equation}
U(\alpha g_1 + \beta g_2)(\boldsymbol{x}_i) = \alpha g_1(\boldsymbol{x}_{i+1}) + \beta g_2(\boldsymbol{x}_{i+1}) = \alpha Ug_1(\boldsymbol{x}_i) + \beta U g_2(\boldsymbol{x}_i).
\end{equation}
Linear map acting on a linear space, though both are infinite-dimensional, raises the spectral problem. In fact, the Koopman operator admits a unique decomposition of singular and regular parts~\citep{reed1972methods,mezic2005spectral}.
\begin{equation}
U = U_s + U_r,
\end{equation}
where $U_s$, $U_r$ are a singular and a regular operator respectively defined on the same manifold $\mathcal{M}$. $U_s$ has pure discrete spectrums while $U_r$ has continuous spectrums. For simplicity this paper consider only $U = U_s$. Spectral problem of Koopman operator reads
\begin{equation} \label{eqn:Koopmanspectrumdef}
U \phi_i(\boldsymbol{x}) = \rho_i \phi_i(\boldsymbol{x}), \quad i = 1, 2, \cdots,
\end{equation}
where $\phi_i(\boldsymbol{x}) : \mathcal{M}\rightarrow \mathbb{R}$ is the Koopman eigenfunction, and $\rho_i \in \boldsymbol{C}$ is the Koopman spectrum. 

It will be clear later, that the spectral problem of a linear system is part of the Koopman spectral problem defined on that linear system. And the linear spectrums of the linearized system provides a subset of Koopman spectrums of the original nonlinear systems. Therefore, discrete Koopman spectrums are common, since fluid systems with bounded domain usually have discrete spectrums. On the other hand, continuous spectrums usually appear in the unbounded fluid system. For instance, B\'enard convection between infinite horizontal planes and the plane Poiseuille flow have continuous spectrum~\citep{grosch1978continuous,drazin1982hydrodynamic}. Therefore, the continuous Koopman spectrums are not uncommon.

The Koopman operator can be defined on continuous differential systems~(\ref{eqn:autonomousnonlinear}) as well. For simplicity, definition of the discretized system~(\ref{eqn:discretesystem}) is presented since the continuous system can always be reasonably discretized. Later, the reader will realize the continuous and discretized systems will share the same eigenfunctions and there is a logarithmic relation between their spectrums.

The infinite dimensional eigenfunctions $\phi_i(\boldsymbol{x})$ provide a set of bases for functional analysis. An observable $g(\boldsymbol{x})$ defined on $\mathcal{M}$ can be expanded by above bases
\begin{equation}\label{eqn:Koopmandec}
g(\boldsymbol{x}) = \sum_i a_i \phi_i(\boldsymbol{x}) + r(\boldsymbol{x}).
\end{equation}
If $\phi(\boldsymbol{x})$s are complete bases of mapping $\mathcal{M}\rightarrow \mathbb{R}$, residue $r(\boldsymbol{x})$ is zero. The evolution of observable $g(\boldsymbol{x})$ is obtained by applying the Koopman operator 
\begin{equation}
g(\boldsymbol{x}_{n+1}) = Ug(\boldsymbol{x}_n) = U \left( \sum_i a_i \phi_i(\boldsymbol{x}_n) + r(\boldsymbol{x}_n) \right) = \sum_i a_i \rho_i \phi_i(\boldsymbol{x}_n) + r(\boldsymbol{x}_{n+1}).
\end{equation}
Moreover,
\begin{equation}\label{eqn:evoluteobservable}
g(\boldsymbol{x}_n) = U^n g(\boldsymbol{x}_0) = \sum_i a_i \rho_i^n \phi_i(\boldsymbol{x}_0) + r(\boldsymbol{x}_n) = \sum_i a_i \rho_i^n  
\end{equation}
by absorbing the constant $\phi_i(\boldsymbol{x}_0)$ to $a_i$ and assuming $\phi_i(\boldsymbol{x})$s are complete.

Koopman spectrums for a general system are usually hard to obtain. However, for linear systems, they are compatible with the linear spectrum theory and can be conveniently constructed. The following sections will introduce Koopman spectrums for various linear systems.

\subsection{Koopman spectrums of LTI systems}\label{sec:KoopmanLTI}

Koopman spectrums for LTI systems are studied in various literature. Some of the results are reviewed here for completeness. 

For a linear time-invariant system
\begin{equation} \label{eqn:cLTI}
\dot{\boldsymbol{x}} = A \boldsymbol{x},
\end{equation}
the discretized form by a constant time interval $\tau$ reads
\begin{equation} \label{eqn:dLTI}
\boldsymbol{x}_{n+1} = e^{\tau A} \boldsymbol{ x}_n = A' \boldsymbol{x}_n\,,
\end{equation}
where matrix exponential $e^{\tau A}$ is defined as~\citep{boyce1992elementary} 
\begin{equation} \label{eqn:matrixexponent}
e^{\tau A} = I + \tau A+ \frac{\tau^2 A^2}{2!} + \cdots + \frac{\tau^n A^n}{n!} + \cdots.
\end{equation}
Matrix exponential $e^{\tau A}$ share the same eigenvectors as matrix $A$, and the spectrums $\rho$ of $e^{\tau A}$ are related to that ($\lambda$) of $A$ by
\begin{equation}\label{eqn:dis-cont}
\rho = e^{\tau \lambda}.
\end{equation}

\subsubsection{Koopman spectrums of the LTI systems}
\label{sec:spectrumlinear}

A linear observable defined by
\begin{equation}
\phi(\boldsymbol{x}) = (\boldsymbol{x}, \boldsymbol{w}),
\end{equation}
is found to be a Koopman eigenfunction~\citep{rowley2009spectral}, since
\begin{equation}
U \phi(\boldsymbol{x}) = (A'\boldsymbol{x}, \boldsymbol{w}) = (\boldsymbol{x}, A'^*\boldsymbol{w}) = (\boldsymbol{x}, \bar{\rho}\boldsymbol{w}) = \rho (\boldsymbol{x}, \boldsymbol{w}) = \rho \phi(\boldsymbol{x}) .
\end{equation}
Here $\boldsymbol{w}$ is the left eigenvector of $A'$ ($A'^*\boldsymbol{w} = \bar{\rho} \boldsymbol{w}$), and $(\cdot, \cdot)$ is the inner-product. $\cdot^*$ is for matrix Hermitian. Therefore, Koopman spectrum $\rho$ equals to the spectrum of linear matrix $A'$. 

For a continuous differential system~(\ref{eqn:cLTI}), the Koopman operator can be defined as
\begin{equation}
U g(\boldsymbol{x}) = g(A \boldsymbol{x}),
\end{equation}
its Koopman spectrum $\lambda$ is related to $\rho$ of the discretized system~(\ref{eqn:dLTI}) by the logarithmic relation~(\ref{eqn:dis-cont}). Besides, the two systems share the same eigenmode. For distinguishing purpose, the spectrum of the discrete system ($\rho$) is called \emph{Koopman multiplier} and the spectrum of the continuous system ($\lambda$) is called \emph{Koopman exponent}.

\subsubsection{Spectrums for nonlinear observables}

In situation where quadratic observable such as $\boldsymbol{x}^TM\boldsymbol{x}$ is examined, a quadratic observable
\begin{equation} \label{eqn:quadobservable}
\phi(\boldsymbol{x}) = (\boldsymbol{x}, \boldsymbol{w}_i)(\boldsymbol{x}, \boldsymbol{w}_j)
\end{equation}
provides the necessary Koopman eigenfunction, since
\begin{equation} \label{eqn:quadob}
U \phi(\boldsymbol{x}) = \phi(A'\boldsymbol{x}) = (A'\boldsymbol{x}, \boldsymbol{w}_i)(A'\boldsymbol{x}, \boldsymbol{w}_j) = \rho_i \rho_j (\boldsymbol{x}, w_i)(\boldsymbol{x}, w_j)= \rho_i \rho_j \phi(\boldsymbol{x}).
\end{equation}
Therefore, if $\rho_i$, $\rho_j$ are spectrums of $A'$, $\rho_i\rho_j$ is an Koopman multiplier corresponding to above quadratic observable~\citep{rowley2009spectral}. Or $ \lambda_i+\lambda_j$ is the Koopman exponent~\citep{budivsic2012applied}. The derived spectrum describes the dynamics of the quadratic observable.

For another nonlinear observable $\frac{1}{\boldsymbol{x}}$, the observable
\begin{equation} \label{eqn:ratiobservable}
\phi(\boldsymbol{x}) = \frac{1}{(\boldsymbol{x}, \boldsymbol{w})}
\end{equation}
gives the new eigenfunction, since 
\begin{equation} \label{eqn:ratiob}
U\phi(\boldsymbol{x}) = \phi (A'\boldsymbol{x}) = \frac{1}{(A'\boldsymbol{x}, \boldsymbol{w})} = \frac{1}{\rho} \phi(\boldsymbol{x}),
\end{equation}
if $\boldsymbol{x}_n$ is not orthogonal to $\boldsymbol{w}$. Correspondingly, $\frac{1}{\rho}$ is the Koopman multiplier, or $-\lambda$ is the Koopman exponent.

The generation of a new Koopman spectrum from the old via nonlinear observable is called \emph{proliferation}. Other nonlinear observables can be first expanded by the Taylor series. Koopman eigenfunctions can be similarly constructed and new Koopman spectrums are proliferated accordingly.

\subsection{Koopman spectrums of LTV systems}\label{sec:KoopmanLTV}
%

For a LTV system
\begin{equation} \label{eqn:LTV}
\dot{\boldsymbol{x}} = A(t) \boldsymbol{x},
\end{equation}
with initial condition $\boldsymbol{x}(t_0) = \boldsymbol{x}_0$, $\boldsymbol{x} \in \mathbb{R}^N$ and $A(t) \in \mathbb{R}^{N\times N}\oplus \mathbb{R}$, the spectrums of $A(t)$ are no longer constant. They reflect the instantaneous dynamics but not the global ones. Instead, the spectrums of \textit{fundamental matrix} of the linear dynamic system can reflect the global dynamics. A matrix $\Psi(t) \in \mathbb{R}^{N\times N}\oplus \mathbb{R}$ is called \textit{fundamental matrix} of LTV system if each column of $\Psi(t)$ satisfy Eqn.~(\ref{eqn:LTV}) and $\Psi(t)$ is not singular~\citep{boyce1992elementary}. The fundamental matrix provides useful dynamics and stability information~\citep{wu1974note}. For example, the fundamental matrix provides global stability indicator of the LTV system but not $A(t)$. \citet{wu1974note,zubov1962mathematical} gave examples when even $A(t)$ had all constant negative eigenvalues, the system was unstable, or the LTV system was stable even $A(t)$ had positive eigenvalues.

Since $\Psi(t)$ is non-singular, a particular fundamental matrix $\Phi(t)$ is defined
\begin{equation}\label{eqn:PFM}
\Phi(t,t_0) = \Psi(t) \Psi(t_0)^{-1}.
\end{equation}
$\Phi(t_0,t_0)=I$. It can be tested that $\boldsymbol{x}(t) = \Phi(t,t_0) \boldsymbol{x}_0$ satisfies the LTV system and the initial condition. Therefore, it is the solution of the LTV system.

\subsubsection{Koopman spectrums of LTV systems}

The discretized LTV system~(\ref{eqn:LTV}) is written
\begin{equation} \label{eqn:LTVdiscrete}
\boldsymbol{x}_{n+1} = \Phi(t_{n+1}, t_n) \boldsymbol{x}_n,
\end{equation}
where $\Phi(t_{n+1}, t_n)$ is the matrix from Eqn.~(\ref{eqn:PFM}).

Similarly, Koopman operator acting on the discrete system~(\ref{eqn:LTVdiscrete}) has Koopman spectrum $\rho(t_n)$ and the corresponding Koopman eigenfunction is given by
\begin{equation} \label{eqn:ltveigenfunction}
\phi(\boldsymbol{x}, t_n) = (\boldsymbol{x}, \boldsymbol{w}_n),
\end{equation}
since
\begin{equation} \label{eqn:eigenfun4LTV}
\begin{aligned}
U \phi(\boldsymbol{x}, t_n) &= \phi(\Phi(t_{n+1},t_n)\boldsymbol{x}, t_n) = (\Phi(t_{n+1},t_n)\boldsymbol{x}, \boldsymbol{w}(t_n)) \\ &= (\boldsymbol{x}, \Phi(t_{n+1},t_n)^*\boldsymbol{w}(t_n))= (\boldsymbol{x}, \bar{\rho}(t_n)\boldsymbol{w}(t_n)) = \rho(t_n) \phi(\boldsymbol{x}, t_n).
\end{aligned}
\end{equation}
where $\rho(t_n)$ and $\boldsymbol{w}_n$ are eigenvalue and left eigenvector of matrix $\Phi(t_{n+1},t_n)$. The time-dependent Koopman spectrum $\rho(t_n)$ provides the dynamic information, such as the growth or decay rate and frequency from time $t_n$ to $t_{n+1}$.

After obtaining the transient spectrum and eigenspace, the one-step evolution of dynamics is achieved by Koopman decomposition
\begin{equation}
g(\boldsymbol{x}_{n+1}) = U g(\boldsymbol{x}_n) = U \left(\sum_{i=1}^{\infty} a_i(t_n) \phi_i(\boldsymbol{x}_n, t_n)\right) = \sum^{\infty}_{i=1} a_i(t_n) \rho_{i}(t_n)\phi_i(\boldsymbol{x}_n, t_n) ,
\end{equation}
where the observable $g$ is decomposed by the Koopman eigenfunction at time $t_n$
\begin{equation} \label{eqn:KoopmaninitdecLTV}
g(\boldsymbol{x}) = \sum_{i=1}^{\infty} a_i(t_n) \phi_i(\boldsymbol{x},t_n).
\end{equation}
Eigenfunction defined in Eqn.~(\ref{eqn:ltveigenfunction}) depends on time $t$, therefore, the Koopman decomposition, the coefficient $a_i(t)$s depend on time. Hence, it is generally not straightforward for a LTV systems to advance the dynamics via Koopman decomposition at all time instances as previous simple LTI systems. Except in a special case when $A(t)$ is only a function of the state or there is a one-to-one correspondence between $t$ and $\boldsymbol{x}$, for instance, $x = x(t)$ and there is an implicit relation $t = t(x)$, so
\begin{equation}
A(t) = B(\boldsymbol{x}).
\end{equation}
In such a situation, eigenfunction $\phi$ is only function of state $\boldsymbol{x}$, so decomposition~(\ref{eqn:KoopmaninitdecLTV}) is independent on $t$.

The Koopman exponent for the underlined continuous system is then defined by
\begin{equation}
\lambda( t_n) = \frac{\ln \rho(t_{n+1}, t_n)}{t_{n+1}- t_n}.
\end{equation}
$t_{n+1}- t_n\rightarrow 0$ provides the transient spectrum at $t_n$. Two special cases exist.

\subsubsection{$\tau = t- t_0$ goes to $\infty$}

A special case is $\tau \rightarrow \infty$, and if the limit exists,
\begin{equation} \label{eqn:spectruminfty}
\lambda(t_0) = \lim_{\tau \rightarrow \infty}\frac{\ln \rho(t_0+\tau, t_0)}{\tau}.
\end{equation}
$\lambda(t_0)$ decides the exponential stability (ES) of the dynamic system. If all $\text{Re}(\lambda(t_0)) \leq 0$, the system is stable, otherwise unstable. The ES can derive the well known asymptotic stable (AS). If $\lambda(t_0)$ is further independent on $t_0$, the uniform asymptotic stable (UAS) is obtained~\citep{zhou2016asymptotic,antsaklis2007linear}, which is a uniform global stability indicator.

\subsubsection{$A(t)$ is periodic}

Another special case is when $A(t)$ is periodic.
\begin{equation} \label{eqn:periodic0}
\dot{\boldsymbol{x}} = A(t) \boldsymbol{x}, \quad A(t+T) = A(t) \text{ and } \boldsymbol{x}(t_0) = \boldsymbol{x}_0.
\end{equation}
This is a Floquet system and its solution is described by the Floquet theory, see Appendix~\ref{sec:periodicLTV}. This type of LTV system is usually obtained by linearizing a system around a periodic solution.

For these Floquet systems, Floquet exponents are the Koopman exponents, corresponding Koopman eigenfunctions are time-periodic functions $\phi_i(\boldsymbol{x},t)$, see Eqn.~\ref{eqn:eigenfunctionfloquet}.

Further transferring periodic function $\phi_{il}(\boldsymbol{x},t)$ to Fourier wavespace, see Eqn.~\ref{eqn:eigenfunctionfloquetfourier}, another Koopman eigenfunction is obtained. The Koopman spectrum is then
\begin{equation}
\lambda_{il} = \mu_i + j l \omega
\end{equation}
Here $\mu_i$ is Koopman exponent. $\omega = \frac{2\pi}{T}$ is the natural frequency, and $i, l \in \mathbb{N}$. Therefore, periodic systems have time-invariant Koopman spectrums.

\section{Local Koopman spectral problem for nonlinear systems} \label{sec:sepctrumnonlinear}


For a nonlinear autonomous dynamic system
\begin{equation}\label{eqn:autonomousnonlinear}
\dot{\boldsymbol{x}} = \boldsymbol{f} (\boldsymbol{x}), \quad \boldsymbol{x}(0) = \boldsymbol{x}_0,
\end{equation}
global Koopman spectrum (independent of $\boldsymbol{x}$) defined in Eqn.~(\ref{eqn:Koopmanspectrumdef}) describes the global dynamic characteristics. As mentioned before, global spectrums are widely used to determine if an system is ergodic. If all fixed points of the defined Koopman operator $U$ are constant functions~\citep[see, chap. 4.2]{lasota2013chaos}, then the system is ergodic. Therefore, for an ergodic system $\lambda=1$ (the fix point) is the spectrum of Koopman operator, and the corresponding eigenfunction is everywhere constant.

However, difficulties persist when applying above global Koopman spectrums to dynamics analysis. Firstly, it is hard to prove such Koopman spectrums exist for a general nonlinear system. Secondly, even if they exist, there lacks an algorithm to compute them. More importantly, some important spectrums characterizing the dynamics are not included in the global Koopman spectrums. For example, in a Hopf bifurcation process, the unstable spectrums (from linear stability analysis) around the unstable equilibrium are not. Otherwise, we could construct a particular initial condition, where only components associated with the growing spectrums are nontrivial. From this particular initial condition, the dynamics exponentially grow without bound, which contradicts to the physical system.
 
The above difficulties can be solved by defining local Koopman spectrums for nonlinear dynamics.
\begin{equation} \label{eqn:Koopmanspectrumlocal}
U\phi(\boldsymbol{x}) = \rho \phi(\boldsymbol{x}), \quad \boldsymbol{x} \in D(\boldsymbol{x})
\end{equation}
Here $D(\boldsymbol{x})$ is an open domain covering $\boldsymbol{x}$. $\lambda$ is the local Koopman spectrum and $\phi(\boldsymbol{x})$ is the corresponding eigenfunction. The local spectrum is backward compatible with the global spectrum, since $D(\boldsymbol{x}) $ can be the whole manifold $\mathcal{M}$,  for instance, LTI systems admit global spectrums.

\subsection{Semigroup Koopman operator}

As spectrums vary from state to state, for a discretized system, they must depend on the state as well as the time step of the discretization. Therefore, a time-parameterized Koopman operator, also called a semigroup operator, is introduced.

A \emph{semigroup operator} $T^{t}$ is defined on the system~(\ref{eqn:autonomousnonlinear}) to evolute the dynamics
\begin{equation}
\boldsymbol{x}(t) = T^t \boldsymbol{x}_0
\end{equation}
for $\forall t \in [0, \infty)$. $T^t$ is the semigroup operator which satisfy the following two requirements
\begin{itemize}
\item the initial condition: $T^0 = I,$
\item time-translation invariance: $T^{t+s} = T^t T^s$
\end{itemize}
The first requirement is easy to check. For the second condition, since the system is autonomous
\begin{equation}
T^tT^s \boldsymbol{x}_0 = T^t(T^s \boldsymbol{x}_0) = T^t(\boldsymbol{x}(s)) = \boldsymbol{x}(t+s) = T^{t+s}\boldsymbol{x}_0.
\end{equation}

The induced semigroup Koopman operator is defined such that
\begin{equation}
U^t g(\boldsymbol{x}) = g(T^t \boldsymbol{x}).
\end{equation}
It can be checked the defined operator $U^t$ is linear and satisfies the two semigroup operator requirements. Therefore, it is also called semigroup Koopman operator. From now on, the suffix $0$ on the initial condition will be dropped, since it works for the initial condition at arbitrary time.

$t\in [0, \infty)$ is the parameter for the evolution interval. In the following application, $t$ is chosen a fixed value of $\tau$; that is, the system~(\ref{eqn:autonomousnonlinear}) is discretized by a constant time interval $\tau$. For simplicity, we will drop the parameter $\tau$ and write the semigroup Koopman operator in the usual way $U$.

\subsection{Hierarchy of local Koopman spectrums for nonlinear dynamics}

Computing Koopman spectrums for a general nonlinear system is a challenging task. However, if augmented with extra information, such as the real trajectory, we could resolve the hierarchy of Koopman eigenspaces for nonlinear dynamics. To see it, the dynamics are decomposed into a base (a real trajectory) and infinitesimal perturbation on top of it. The spectrums are computed for the base and perturbation part separately.

\subsubsection{The spectrums of base}
For a simple base flow, Koopman spectrums can be obtained analytically. For instance, for a nonlinear system at a fixed equilibrium state, the base flow is the equilibrium state. The dynamics of this base flow are merely given by spectrum of $\rho=1$ or $\lambda=0$, for discretize or continuous system respectively.

Another example of the base is the periodic solution embedded in a limit cycle solution of a nonlinear system. Then the Fourier spectrums $\pm jnw$ ($j=\sqrt{-1}$, $n \in \mathbb{N}$, $w= \frac{2\pi}{T}$) provide the base Koopman spectrum.

For general dynamics, baseflow can be a \textit{real trajectory} $\boldsymbol{x}(t)$. Then numerical techniques such as the DMD algorithm can be applied for its spectrums. 

The base should be a real trajectory of the system, otherwise, fake spectrums may be generated. For instance, if the system is ergodic and the sampling is long enough, the mean is the fixed point of the unitary operator $U$, whose spectrums $\lambda=0$ is indeed included in the Koopman spectrums~\citep{mezic2005spectral}. Therefore, mean flow subtraction will not change the spectrums. However,~\citet{rowley2009spectral} pointed out mean subtraction introduces fake Fourier spectrums. The failure is because the mean flow is not a real solution for the considered dynamic system.

Since subtraction of real trajectory won't change Koopman spectrums, a procedure to filter out base spectrums and to focus on perturbation spectrums can be designed by subtracting a piece of real trajectory from another real trajectory. For instance, if real trajectory $X=\{\boldsymbol{x}_1, \cdots, \boldsymbol{x}_{N+1}\}$ of a dynamic system is given, the perturbation $X'=\{\boldsymbol{x}'_1, \cdots, \boldsymbol{x}'_{N}\}$ can be constructed by
\begin{equation}
\boldsymbol{x}'_i = \boldsymbol{x}_{i+1} - \boldsymbol{x}_i.
\end{equation}
DMD algorithm on the perturbation data $X'$ will compute the perturbation spectrums. Though this procedure does not provide new information compared to applying DMD on the original data sequence $X$, it provides a method to subtract the base dynamics from the data and to focus on the spectrums of perturbation.

\subsubsection{The spectrums of perturbation}

The spectrum of perturbation can be obtained by considering the nonlinear perturbation equation
\begin{equation}\label{eqn:LTIperturbation}
\dot{\boldsymbol{x}}' = A(t)\boldsymbol{ x}' + \mathcal{N}(\boldsymbol{x}').
\end{equation}
$\boldsymbol{x}'$ is the small magnitude perturbation on top of the base flow and
\begin{equation}
\boldsymbol{x} = \boldsymbol{X} + \boldsymbol{x}'.
\end{equation}
Here $\boldsymbol{X}$ is the base flow satisfying $\dot{\boldsymbol{X}} = f(\boldsymbol{X})$. 

For a steady equilibrium solution $\boldsymbol{X}$, $A=\frac{\partial f}{\partial \boldsymbol{x}}\big|_{\boldsymbol{X}}$ is a constant matrix. For an unsteady base $\boldsymbol{X}(t)$, $A(t)=\frac{\partial f}{\partial \boldsymbol{x}}\big|_{\boldsymbol{X}(t)}$ is time-dependent. $\mathcal{N}(\boldsymbol{x}')$ is the nonlinear interaction of perturbation. The spectrums of the underlined linear systems considered in \S~\ref{sec:KoopmanLTI} or \S~\ref{sec:KoopmanLTV} provide a subset of Koopman spectrums as $\boldsymbol{x}'\rightarrow 0$, viz, the local spectrum. 

Moreover, the nonlinear perturbation Eqn.~(\ref{eqn:LTIperturbation}) shows $\boldsymbol{x}'$ is a nonlinear function of itself, therefore, a nonlinear observable. Note the proliferation applies to nonlinear systems as well, since if $\phi_i$, $\phi_j$ are eigenfunction of $U$,
\begin{equation}
U(\phi_i \phi_j)(x) = (\phi_i\phi_j)(T^t x) = \phi_i(T^tx) \phi_j(T^tx) = \rho_i \rho_j (\phi_i\phi_j)(x).
\end{equation}

Assuming the nonlinear term $\mathcal{N}(\boldsymbol{x}')$ can be expanded by Taylor series, such that
\begin{equation}
\mathcal{N}(\boldsymbol{x}') = \mathcal{N}_2 \boldsymbol{x}'^2 + \mathcal{N}_3 \boldsymbol{x}'^3 + \cdots.
\end{equation}
Here $\mathcal{N}_i \boldsymbol{x}'^i$ is the symbolic notation, which should be understood as tensor product between order $i$ tensor $\boldsymbol{x}^i$ and order $i+1$ tensor $\mathcal{N}_i $. For each nonlinear observable $\mathcal{N}_i \boldsymbol{x}'^{i}$, the proliferation rule applies. Moreover, the proliferation rule is applied \emph{recursively}. Since from the nonlinear perturbation equations the nonlinear relation between $\boldsymbol{x}'$ and $\mathcal{N}_i \boldsymbol{x}'^{i}$ is implicit, the process can continue after applying proliferation rule to obtain Koopman spectrum for nonlinear observables. In contrast, a nonlinear observable on a linear dynamic system discussed in section~\ref{sec:KoopmanLTI} only admits one proliferation.

\subsection{Nonlinearity and spectrum patterns}

Since nonlinear dynamics can recursively generate new spectrums, and each of them represents particular dynamics, therefore, it is useful to consider these proliferated spectrums.

\subsubsection{Self interaction}

\begin{figure}
\centering
\begin{subfigure}[b]{0.325\linewidth}
\includegraphics[width=1.0\textwidth]{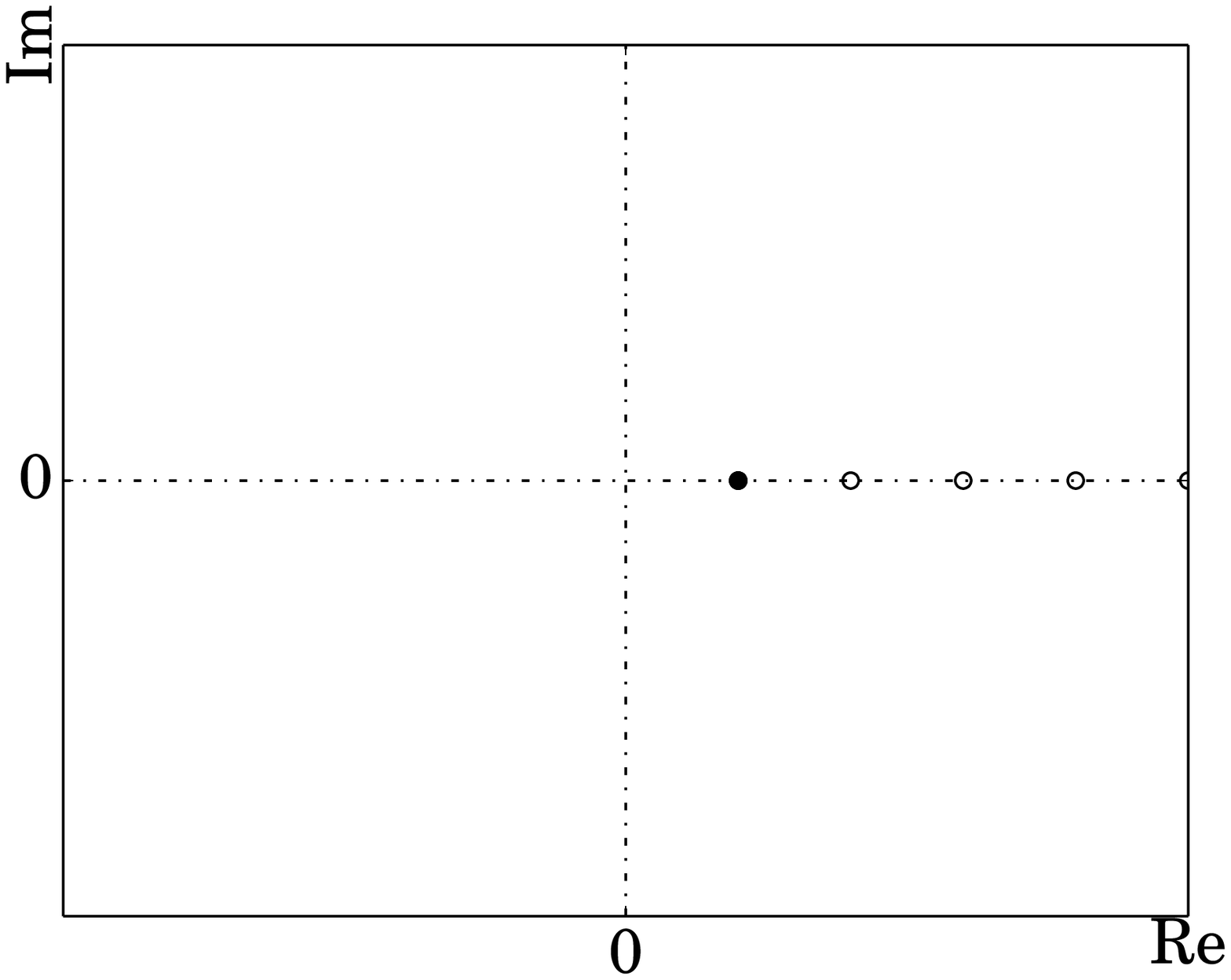}
\caption{$\lambda>0$ \label{fig:selfreal} }
\end{subfigure}
\begin{subfigure}[b]{0.325\linewidth}
\includegraphics[width=1.0\textwidth]{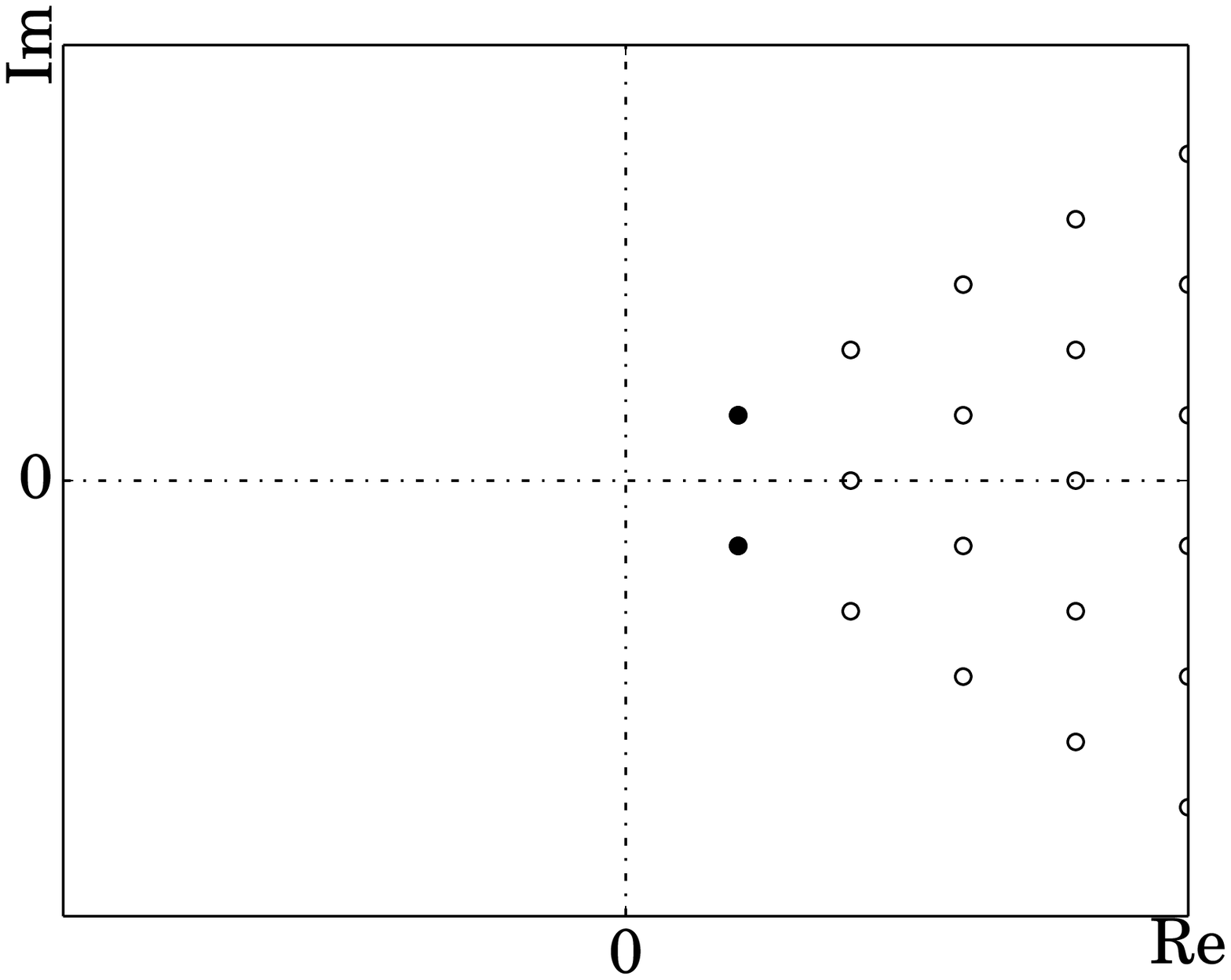}
\caption{$\text{Im}(\lambda)\neq 0$ \label{fig:selfcomplex} }
\end{subfigure}
\begin{subfigure}[b]{0.325\linewidth}
\includegraphics[width=1.0\textwidth]{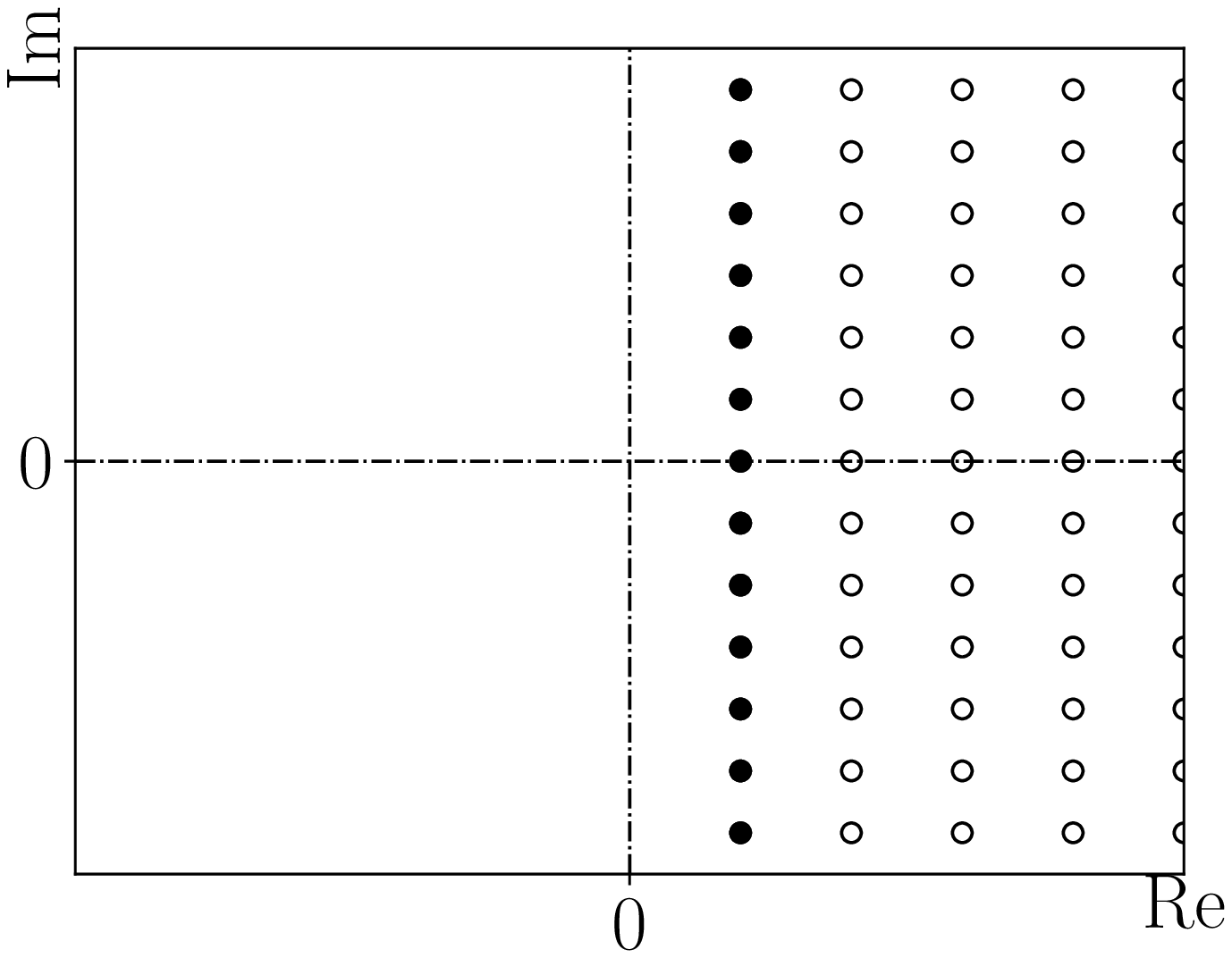}
\caption{limit cycle system \label{fig:selffloquet} }
\end{subfigure}
\caption{Self interaction of Koopman spectrums. The filled circles show the original linear spectrums, high-order spectrums are shown by hollow circles.} \label{fig:selfinteract}
\end{figure}

For example, if $\lambda$ is a real Koopman exponent for the underlined linear system of 
\begin{equation} \label{eqn:quaddyn}
\dot{\boldsymbol{x}} = A\boldsymbol{x} + \mathcal{N}_2\boldsymbol{x}^2, 
\end{equation}
by the recursive proliferation rule, the following are also Koopman spectrums
\begin{equation}
\lambda, \lambda + \lambda, (\lambda+\lambda)+\lambda, ((\lambda+\lambda)+\lambda)+\lambda, \cdots.
\end{equation}
These spectrums fall on a line and are shown in figure~\ref{fig:selfreal}.

If complex conjugate spectrums $\lambda$, $\bar{\lambda}$ are the spectrums of the underlined linear system~(\ref{eqn:quaddyn}), the Koopman spectrums proliferated by them fall on a triad-chain shown in figure~\ref{fig:selfcomplex}.

One more example is the following time-dependent system
\begin{equation*}
\dot{\boldsymbol{x}} = A(t)\boldsymbol{x} + \mathcal{N}_2\boldsymbol{x}^2, 
\end{equation*}
and $A(t)$ is periodic. From Floquet theory and discussion in \S~\ref{sec:periodicLTVprove} the linear part provide spectrums $\lambda \pm jmw$, where $j=\sqrt{-1}$ and $m$ is an integer. By applying nonlinear proliferation, the lattice distribution of Koopman spectrums are obtained
\begin{equation}
n\lambda \pm jmw,
\end{equation}
see figure~\ref{fig:selffloquet}. Here $n$ is a positive integer. \citet{bagheri2013koopman} came to the same conclusion by analytically computing spectrums of Frobenious-Perron operator, the adjoint operator of Koopman, via the trace formula on K\'arm\'an vortex flow.

All the above self-interacted spectrums share the same positiveness or negativeness of the real part as their parental spectrums $\lambda$ and $\bar{\lambda}$. Therefore, $\mathcal{N}_2\boldsymbol{x}^2$ term does not change the stability of the dynamic system in the asymptotic case, neither other $\mathcal{N}_i \boldsymbol{x}'^i$ nonlinearity. In this sense, the spectrums of the linearized perturbation equation determine the stability of nonlinear perturbation.

\subsubsection{Cross interaction}

\begin{figure}
\centering
\begin{subfigure}[b]{0.325\linewidth}
\includegraphics[width=1.0\textwidth]{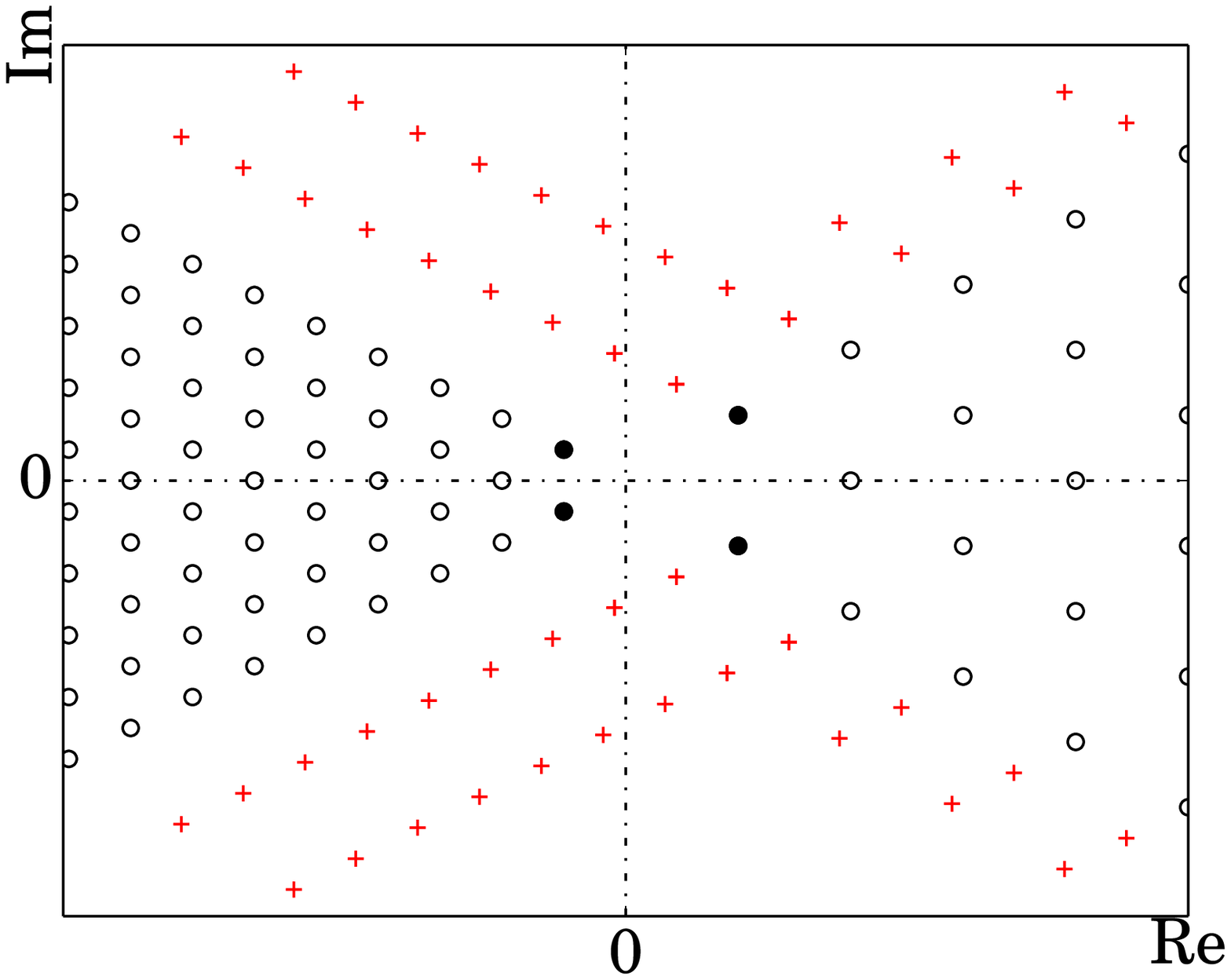}
\caption{ $\text{Re}(\lambda_1+\lambda_2)>0$ \label{fig:crosspos} }
\end{subfigure}
\begin{subfigure}[b]{0.325\linewidth}
\includegraphics[width=1.0\textwidth]{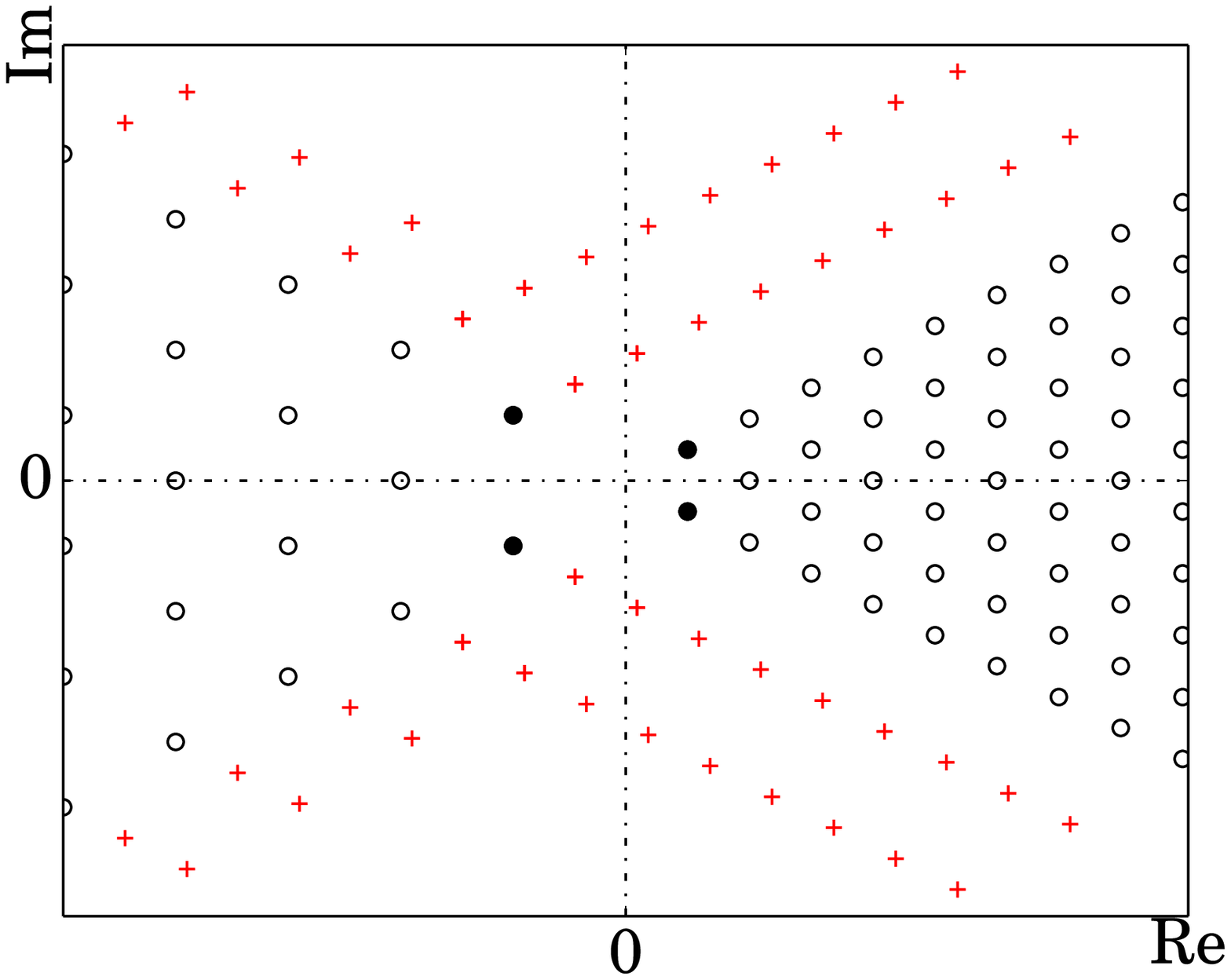}
\caption{ $\text{Re}(\lambda_1+\lambda_2)<0$ \label{fig:crossneg} }
\end{subfigure}
\caption{Cross interaction of Koopman spectrums. The filled circles show the original spectrums, their high order derived spectrums are shown by the hollow circle. Part of the cross interaction spectrum are shown by the red crosses.} \label{fig:crossspectrum}
\end{figure}

Besides above self-interaction between one mode and its complex conjugate, modes interaction can happen between different complex conjugate pairs. The cross interaction will then generate new spectrums as well. For instance, if two pairs of Koopman spectrums both have positive or negative real part, then the derived spectrum should have the same positiveness or negativeness. If they are of different sign, either $\text{Re}(\lambda_1+\lambda_2)>0$ or $\text{Re}(\lambda_1+\lambda_2)<0$ can excite some unstable modes illustrated by figure~\ref{fig:crossspectrum}.

\subsubsection{Rational nonlinear terms}

\begin{figure}
\centering
\begin{subfigure}[b]{0.325\linewidth}
\includegraphics[width=1.0\textwidth]{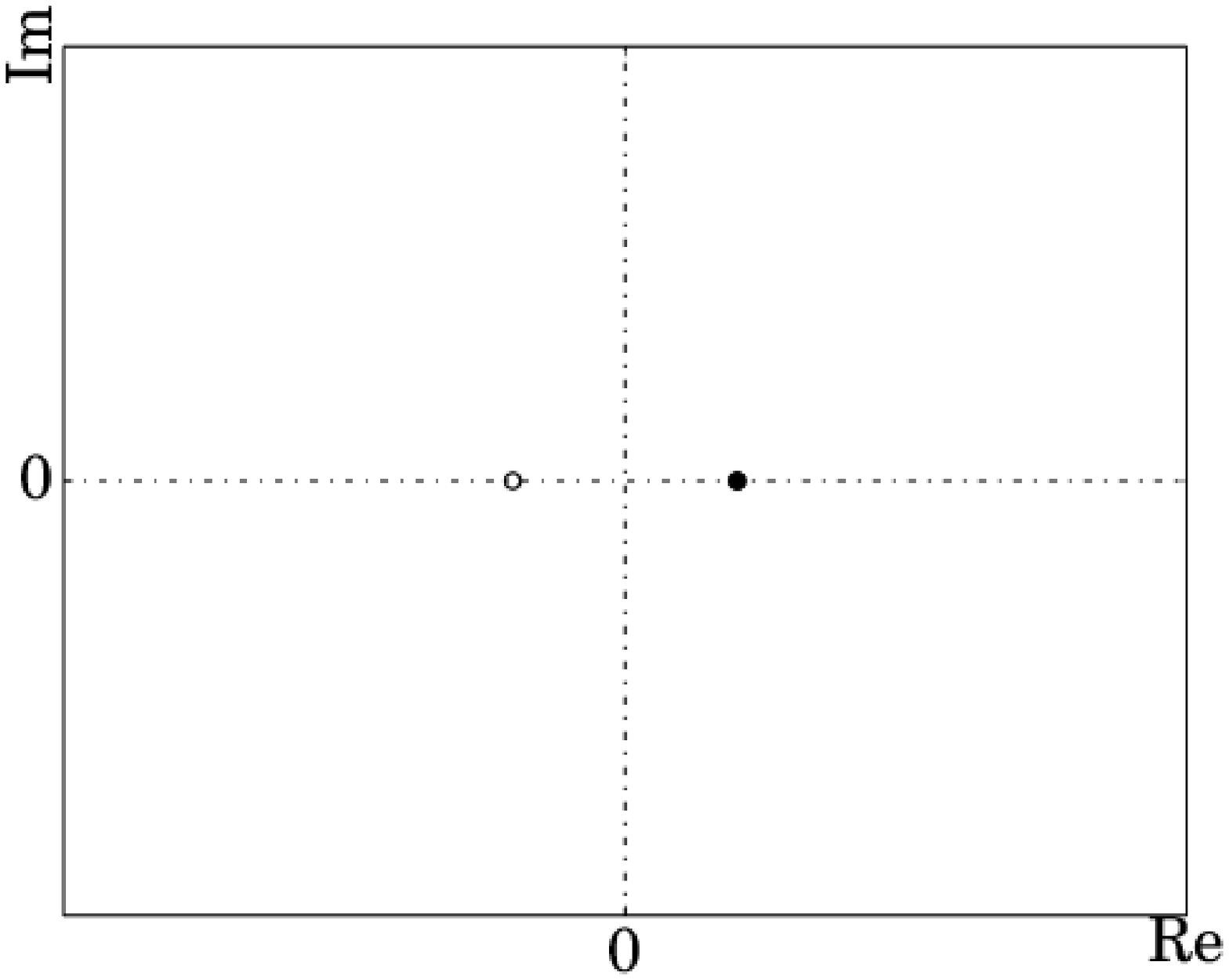}
\caption{ $\lambda > 0 $ }
\end{subfigure}
\begin{subfigure}[b]{0.325\linewidth}
\includegraphics[width=1.0\textwidth]{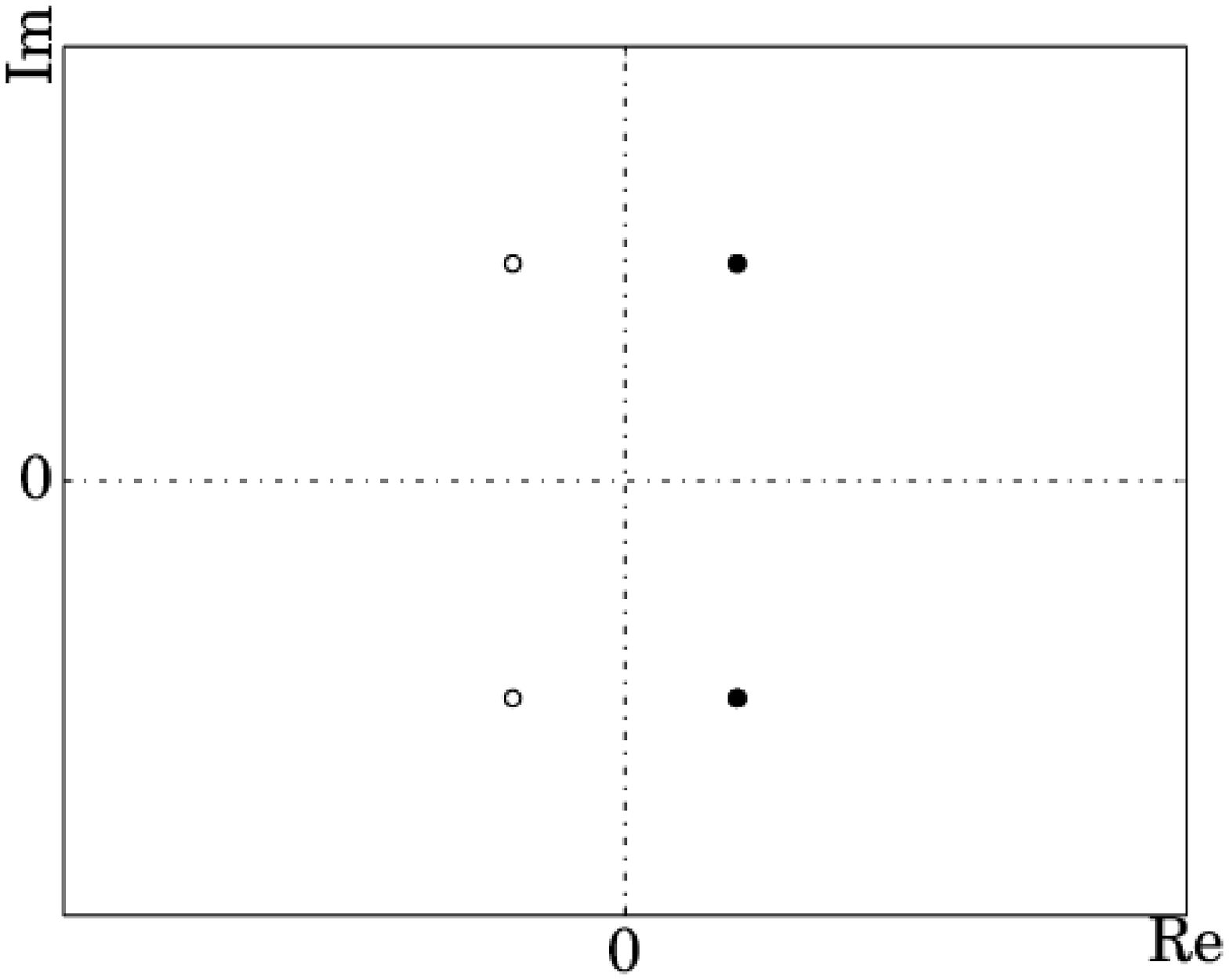}
\caption{ $\text{Re}(\lambda)>0$ and $\text{Im}(\lambda)\neq 0$}
\end{subfigure}
\caption{Self interaction for term $\frac{1}{\boldsymbol{x}}$.} \label{fig:selfrational}
\end{figure}

For the rational nonlinearity $\frac{1}{\boldsymbol{x}}$, the Koopman spectrums are shown in figure~\ref{fig:selfrational}, which indicates there are always a growing and a decaying mode at the neighborhood of $\boldsymbol{x}$.

\subsubsection{Fluid dynamic system: a nonlinear dynamic system example}

For a fluids dynamic system, nonlinearity usually comes from the convection, written as $u_i \frac{\partial u_j}{\partial x_i}$. They are second-order nonlinear terms. Possible spectral patterns are already shown in figure~\ref{fig:selfinteract} and figure~\ref{fig:crossspectrum}. Their corresponding modes, especially those unstable ones, may develop significant flow patterns, making complicated flow phenomena. Therefore, the recursively proliferated spectrums and modes provide an efficient device to describe the nonlinear interaction in fluid systems.

Other than the infinite-dimensional Koopman eigenspaces, there are other descriptions for nonlinear interactions. They were discovered and documented by many authors, for example, resonant wave interaction by~\citet[see \S 51]{drazin1982hydrodynamic} and three-wave interaction by~\citet[see \S 5.4]{schmid2012stability} and others but not listed here. However, it will be clear in part 2 of this paper, Koopman eigendecomposition has advantages over other approaches.

\subsection{Spectral theory and the continuity of local Koopman spectrums}
\label{sec:localkoopmanspectrumcontinuity}


To discuss the existence of local Koopman spectrums, let us introduce some basic definitions from operator theory.

A \emph{Banach space} is a \emph{complete} \emph{normed space}.

A \emph{Hilbert space} is a \emph{complete} \emph{inner product} space.

A \emph{bounded operator} from a normed linear space $\langle V_1, ||\cdot||_1 \rangle$ to a normed linear space $\langle V_2, ||\cdot||_2 \rangle$ is a map $T$, which satisfies
\begin{enumerate}
\item $T(\alpha v + \beta w) = \alpha T(v) + \beta T(w), \quad (\forall v, w \in V_1 \text{ and } \forall \alpha, \beta \in \mathbb{R} \text{ or } \mathbb{C}$), 
\item Exists some $C \ge 0$, $||T (v)||_2 \le C ||v||_1, \quad \forall v \in V_1$.
\end{enumerate}
The spectrums of a bounded operator are well studied. They are known not empty and bounded for bounded operator~\citep[chap. VI.3]{reed1972methods}. On the contrary, the spectrums for the unbounded operator largely remain undetermined. For the fluid systems that are interested in this work, the spatial differentiation results in unbounded operator~\citep[chap. VIII]{reed1972methods}.

The local Koopman spectrum is defined in a manner to include all dynamics relevant spectrums. However, to the best knowledge of the authors, no precedent works dealt with the local spectral problem for operators. Therefore, we don't have a rigorous existence theorem. However, there are several signs for their existence. Firstly, the local spectrums are compatible with the global spectrums in the previous literature. So if the latter exist, they provide a subset of the local spectrums. Secondly, the local Koopman spectrums were computed by various authors. For instance, the Koopman spectrum of periodic systems studied by various authors~\citep{mezic2005spectral,bagheri2013koopman} are the local Koopman spectrum discussed in Appendix~\ref{sec:periodicLTV}. And the spectrums computed by the data-driven algorithm~\citep{rowley2009spectral,schmid2010dynamic} are time-averaged approximations of those local spectrums over a period of time, see \S~\ref{sec:DMDKoopman}. Further, the stability analysis solves linear eigenvalue problems are essential local Koopman spectrum problems.

\subsubsection{The continuity of local Koopman spectrums} \label{sec:continuitylocalKoopman}

Since local Koopman spectrums are state-dependent, it is inconvenient to analyze a nonlinear transient process. Therefore, the relationship between these local spectrums needs to be discovered. In the following section, the \emph{continuity} property is discussed.

The continuity of the local Koopman spectrum can be explained by operator perturbation theory~\citep{kato2013perturbation,reed1978methods}. Operator perturbation theory shows that an operator $U$ which has a bounded spectrum, perturbed by an infinitesimal bounded operator $\epsilon V$ ($V$ is bounded operator) will also have a bounded spectrum. Moreover, the spectrum and eigenfunction of
\begin{equation}
U + \epsilon V
\end{equation}
will not only continuously but also analytically change with parameter $\epsilon$ from the original operator $U$. Here $\epsilon \rightarrow 0$.

The perturbation theory provides a valuable tool to study the continuity of these local spectrums. Let us assume a simple case where the observable function $g(\boldsymbol{x})$ and dynamic system $T^{\tau}$ are analytic and their `derivative' are uniformly continuous. Consider two infinitesimally close state $\boldsymbol{x}$ and $\boldsymbol{x}+\boldsymbol{h}$ in a open domain $D(\boldsymbol{x})$, where $||\boldsymbol{h}||\rightarrow 0$. The Koopman operator at state $\boldsymbol{x}+\boldsymbol{h}$ can be expanded by a perturbed Koopman operator at $\boldsymbol{x}$ such that
\begin{equation}
\begin{aligned}
U g(\boldsymbol{x}+\boldsymbol{h}) &= g ( T^{\tau} (\boldsymbol{x}+\boldsymbol{h})) = g( T^{\tau}\boldsymbol{x} + T'^{\tau}\boldsymbol{h} + O(\boldsymbol{h}^2)) \\
& = g( T^{\tau}\boldsymbol{x}) + g' ( T^{\tau}\boldsymbol{x} )\cdot T'^{\tau}\boldsymbol{h} + O(\boldsymbol{h}^2) \\
& \approx (U^{\tau} + V^{\tau}) g(\boldsymbol{x}).
\end{aligned}
\end{equation}
Here $T'^{\tau}$ is the differential of $T^{\tau}$ at $\boldsymbol{x}$ and $g'$ is the differential of $g$ at $T^{\tau} \boldsymbol{x}$. And $V^{\tau}$ is the perturbation operator defined by
\begin{equation}
V^{\tau} g(\boldsymbol{h}) \equiv g'(T'^{\tau}(\boldsymbol{h}))\cdot T'^{\tau}\boldsymbol{h}.
\end{equation}
In the condition that $g$ and $T^{\tau}$ is uniformly continuous differentiable, the perturbation operator is bounded
\begin{equation}
||V^{\tau} g(\boldsymbol{h})|| = || g' ( T^{\tau}\boldsymbol{x} )\cdot T'^{\tau}\boldsymbol{h}|| \le C ||\boldsymbol{h}||.
\end{equation}
Further, for any given small threshold $\epsilon$, let $||\boldsymbol{h}||\le \frac{\epsilon}{C}$. Then the norm of the perturbation will be smaller than $\epsilon$
\begin{equation}
||V^{\tau} g(\boldsymbol{h})|| \le \epsilon
\end{equation}
From perturbation theory, the simple (or isolated) local spectrum is not only continuous but also analytic in the open domain $D_{\frac{\epsilon}{C}}(\boldsymbol{x})$. $\frac{\epsilon}{C}$ is the radius of the domain.

The continuity requirement can be satisfied by many systems. For instance, an incompressible flow dynamic system with a fixed boundary is uniformly continuous differentiable. Further, if the observable function is a continuously differentiable function of state $\boldsymbol{x}$, Koopman operator then has continuous and analytical eigenvalues and eigenfunctions with respect to states.

In the case that eigenvalues and eigenfunctions are continuous, the open domain $D(\boldsymbol{x})$ can be `glued' piece-by-piece, extending the local spectrum not only continuously but also analytically to whole manifold. Therefore, global eigen-relation of the Koopman operator on the whole manifold holds
\begin{equation} \label{eqn:globalKoopmandecnonlinear}
U\phi_i(\boldsymbol{x}) = \rho_i(\boldsymbol{x}) \phi_i(\boldsymbol{x}), \quad \boldsymbol{x} \in \mathcal{M}.
\end{equation}
$\rho_i(\boldsymbol{x})$ is the analytical state-dependentent spectrum, and $\phi_i(\boldsymbol{x})$ is the analytical eigenfunction in manifold $\mathcal{M}$. Koopman decomposition of an continuous observable $g(\boldsymbol{x})$ then reads
\begin{equation} \label{eqn:nonlinearcontinousspectrum}
g(\boldsymbol{x}) = \sum_{i=0}^{\infty} a_i \phi_i(\boldsymbol{x}), \quad \boldsymbol{x} \in \mathcal{M}.
\end{equation}
A point to note here is that the decomposition coefficient $a_i$ are \emph{state independent}. Therefore, for vector observables, Koopman modes are \emph{state independent}.

The dynamics of the observable is obtained by applying the Koopman operator
\begin{equation} \label{eqn:evoKoopmannonlinear}
g(\boldsymbol{x}_n) = U^n g(\boldsymbol{x}_0) = U^n \sum_{i=0}^{\infty} a_i \phi_i(\boldsymbol{x}_0) = \sum_{i=0}^{\infty} a_i \phi_i(\boldsymbol{x}_0)\prod_{k=0}^{n-1} \rho_{ik},
\end{equation}
where $\rho_{ik}=\rho_i(\boldsymbol{x}_k)$ are the local spectrums. The dynamics are obtained by using the globally analytical Koopman spectrums and eigenfunctions.

\subsubsection{Discontinuity of local spectrums and state-dependent modes}

\begin{figure}
\centering
\begin{subfigure}[b]{0.495\linewidth}
\centering
\includegraphics[width=0.3\textwidth]{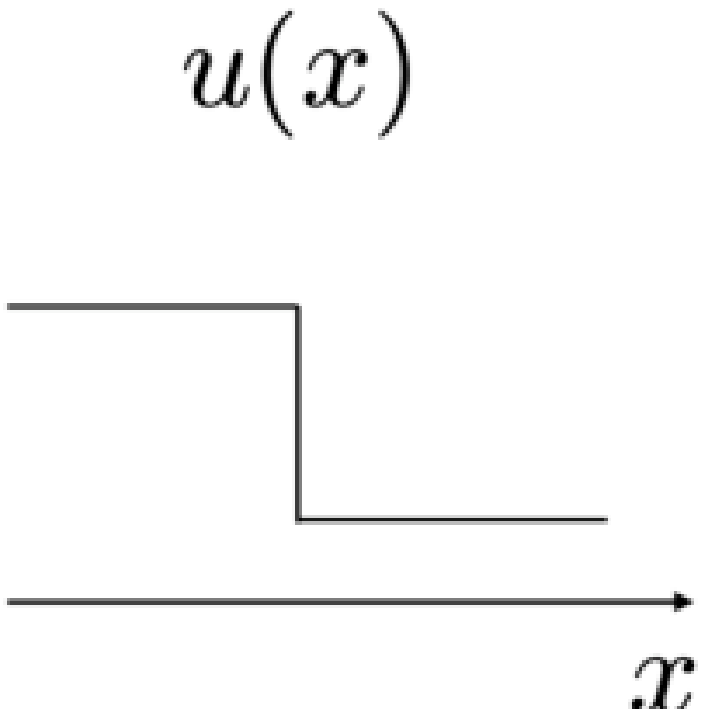}
\caption{shock wave}
\end{subfigure}
\begin{subfigure}[b]{0.495\linewidth}
\centering
\includegraphics[width=0.22\textwidth]{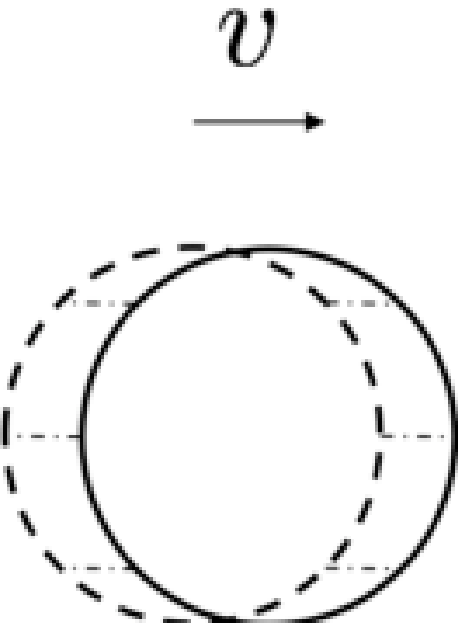}
\caption{moving body}
\end{subfigure}
\caption{Discontinuity for unbounded operators. (a) Shock wave. (b) Moving body.} \label{fig:unboundedKoopman}
\end{figure}

Unfortunately, Koopman operator defined on Navier-Stokes equation may not always be bounded. The differential operator $\frac{\partial }{\partial x}$ can incur $\infty$ at the spatial discontinuity. Figure~\ref{fig:unboundedKoopman} shows two possible examples. The first one is a shockwave problem in a hyperbolic system. The second example is a moving solid in the incompressible flow. As the solid moves, differentiation at the position initially occupied by solid but filled with fluids later (or vise versa) incurs undetermined differentiation.

For above situations, two issues arise for Koopman spectral problem. The first one is the existence of such spectrums, and the second is the continuity of it. The spectrum for unbounded operator, such as the compact operator~\citep[see, chap. VI.5]{reed1972methods} may provide some insight to the first one. Unfortunately, even if these discontinuous local spectrum exist, the continuity relation~(\ref{eqn:globalKoopmandecnonlinear}) may not hold for whole manifold, since these eigenfunctions $\phi_i(\boldsymbol{x})$ are no longer continuous. But the local spectrum (if exist) may be still useful to study the transient dynamics in a local manner
\begin{equation}
g(\boldsymbol{x}) = \sum_{i=0}^{\infty} a_i(\boldsymbol{x}) \phi_i(\boldsymbol{x}), \quad \boldsymbol{x} \in D(\boldsymbol{x}).
\end{equation}

\section{Koopman modes for linear systems}


In many applications, the observable is a vector valued function defined on system state, such that
\begin{equation}
\boldsymbol{g}(\boldsymbol{x}) = \left[ \begin{array}{c} g_1(\boldsymbol{x}) \\ \vdots \\ g_n(\boldsymbol{x}) \end{array} \right].
\end{equation}
Koopman decomposition for this observable is obtained by expanding each component of $\boldsymbol{g}$ by the Koopman eigenfunction $\phi_i(\boldsymbol{x})$, see~(\ref{eqn:globalKoopmandecnonlinear})
\begin{equation}
\boldsymbol{g}(\boldsymbol{x}) \approx \sum \boldsymbol{v}_i \phi_i( \boldsymbol{x} )
\end{equation}
Here $\boldsymbol{v}_i$ is the \textit{Koopman mode}. Each Koopman mode is a collection of component which has the same dynamics, that is, the same growth rate and frequency. They provide rich information of the given dynamic system, especially when the full-state observable $\boldsymbol{x}$ is investigated. In the following discussion, Koopman modes for the full-state observable $\boldsymbol{x}$ are considered.

As mentioned earlier, a remarkable feature of Koopman modes is that they are state-independent for LTI systems, periodic LTV systems. It is also state-independent for the nonlinear autonomous systems under suitable conditions.


\subsection{Example 1. Koopman modes for LTI systems}

Consider a dynamic system 
\begin{equation} \label{eqn:vectorlti}
\boldsymbol{x}_{n+1} = A \boldsymbol{x}_n, \quad A\in \mathbb{R}^{n\times n}.
\end{equation}
Let $A$ diagonalizable and $A = V\Lambda V^{-1}$, and $\lambda_i$, $\boldsymbol{r}_i$, $\boldsymbol{l}_i$ are the eigenvalue, right and left eigenvector. The observable $\boldsymbol{x}$ is decomposed by
\begin{equation}
\boldsymbol{x} = V V^{-1}\boldsymbol{x} = \left( \sum_{i=1}^n \boldsymbol{r}_i \boldsymbol{l}_i^H\right) \boldsymbol{x} = \sum_{i=1}^n \boldsymbol{r}_i \boldsymbol{l}^H_i \boldsymbol{x} = \sum_{i=1}^n \boldsymbol{r}_i \phi_i( \boldsymbol{x} ),
\end{equation}
where $\phi(\boldsymbol{x}_i)$ is the Koopman eigenfunction defined in \S~\ref{sec:spectrumlinear}. Therefore, the right eigenvectors of $A$ are the \emph{Koopman modes} for the full-state observable $\boldsymbol{x}$. 

Using Koopman decomposition, the dynamics of state variable $\boldsymbol{x}$ can be evaluated by
\begin{equation}
\boldsymbol{x}_k = U^k \boldsymbol{x}_0 = \sum_{i=1}^n \boldsymbol{r}_i \rho^k_i \phi_i(\boldsymbol{x}_0),
\end{equation}
which equals to the linear theory since
\begin{equation}
\boldsymbol{x}_k = \sum_{i=1}^n \boldsymbol{r}_i \rho^k_i \phi_i(\boldsymbol{x}_0) = V \Lambda^k V^{-1}\boldsymbol{x}_0 = A^k \boldsymbol{x}_0.
\end{equation}

\subsection{Example 2. Koopman modes for LTV systems}

Similarly, for a LTV system
\begin{equation}
\boldsymbol{x}_{k+1} = A_k \boldsymbol{x}_{k}, \quad A_k \in \mathbb{R}^{n\times n},
\end{equation}
assuming matrix $A_k$s are diagonalizable and $\boldsymbol{r}^k$, $\boldsymbol{l}^k$ are the right and left eigenvectors. The full-state observable $\boldsymbol{x}$ can be decomposed by
\begin{equation}
\boldsymbol{x} = V_k V_k^{-1}\boldsymbol{x} = \left( \sum_{i=1}^n \boldsymbol{r}^k_i (\boldsymbol{l}^{k}_i)^H\right) \boldsymbol{x} = \sum_{i=1}^n \boldsymbol{r}^k_i (\boldsymbol{l}^{k}_i)^H \boldsymbol{x} = \sum_{i=1}^n \boldsymbol{r}_i^k \phi_i( \boldsymbol{x}, t_k ).
\end{equation}
$\phi_i( \boldsymbol{x}, t_k )$ is the eigenfunction~(\ref{eqn:ltvkoopmaneigenfunction}) for the LTV system. Therefore, the time-variant $\boldsymbol{r}^k_i$s are the \emph{Koopman modes} for observable $\boldsymbol{x}$.

\subsection{Example 3. Koopman modes for periodic LTV systems}

For the periodic LTV system, Koopman decomposition using eigenfunction $\phi_i(\boldsymbol{x}, t)$~(\ref{eqn:eigenfunctionfloquet}) is
\begin{equation}
\boldsymbol{x} = Q(t) Q^{-1}(t) \boldsymbol{x} = \sum_{i=1}^{n} \boldsymbol{q}_i(t) \phi_i(\boldsymbol{x}, t).
\end{equation}
Therefore, the columns of $Q(t)$, also known as Floquet modes, are the Koopman modes.

The Floquet modes $\boldsymbol{q}_i(t)$ are time periodic. If transferred to Fourier wavespace, time-invariant Koopman mode $\boldsymbol{q}_{ik}$ is obtained, see Eqn.~(\ref{eqn:invariantFloquetmode}). It corresponds to Koopman eigenfunction~(\ref{eqn:eigenfunctionfloquetfourier}).
$\boldsymbol{q}_{ik}$ is the k-th Fourier component of i-th Floquet mode $\boldsymbol{q}_i(t)$.

\subsection{Koopman modes for nonlinear autonomous systems}

It was mentioned in \S~\ref{sec:continuitylocalKoopman} for `smooth' nonlinear autonomous systems the Koopman modes are independent of the state. However, there are no explicit expressions for them. In part 2 of this paper, asymptotic expansions will be adopted to derive the Koopman modes for nonlinear systems under suitable conditions.

\section{DMD algorithm and local Koopman decomposition} \label{sec:DMDKoopman}

Koopman decomposition can be approximated in a data-driven manner~\citep{rowley2009spectral,schmid2010dynamic}. In this method a set of data $\{\boldsymbol{x}_1, \cdots, \boldsymbol{x}_{m+1} \}$ is first collected. The DMD algorithm is then applied to get the Koopman spectrums and modes.

DMD considers the Koopman operator acting on the system, with a finite-dimensional approximation $K$, such that time-resolved state variables collected from the system will satisfy the following realtion
\begin{equation} \label{eqn:disKoopman}
\boldsymbol{x}_{n+1} = K \boldsymbol{x}_n.
\end{equation}
Here $K\in \mathbb{R}^{n\times n}$ is the discretized Koopman operator which has $n\times n$ dimensions. 

In a recent paper~\citep{zhang2019solving}, the character dynamic information of operator $K$ is effectively extracted by solving a generalized eigenvalue problem (GEV)
\begin{equation}
Y v = \lambda X v,
\end{equation}
where $Y = \left[ \boldsymbol{x}_2\, \cdots \, \boldsymbol{x}_{m+1} \right]$ and $X = \left[ \boldsymbol{x}_1\, \cdots \,\boldsymbol{x}_{m} \right]$. This algorithm has the advantage of circumventing the singularity issue occurred in other DMD algorithms, which is valuable for applications such as stability analysis. The GEV provides the Koopman spectrum $\lambda$ and Koopman mode $X \boldsymbol{v}$ since
\begin{equation}
K (X\boldsymbol{v} ) = Y \boldsymbol{v} = \lambda (X\boldsymbol{v}).
\end{equation}

DMD algorithm is a data-driven technique and can only capture dynamics contained in the data. For example, if only periodic snapshots are given, the Fourier spectrum will be resolved and no spectrums for the perturbation will be obtained. 

DMD algorithm can effectively approximate the spectrums and modes of a system which has constant spectrums. It may fail to capture spectrums for a general LTV system or strong nonlinear systems. In fact, the DMD algorithm provides a time averaged approximation of the local Koopman spectrums. Let us use DMD decomposition and local Koopman decomposition (see Eqn.~\ref{eqn:evoKoopmannonlinear}) to approximate the same data
\begin{equation} \label{eqn:dmdvsKoopman}
\boldsymbol{x}_N = \sum_i \boldsymbol{v}_i \rho_i^N = \sum_i \boldsymbol{v}_i \Pi_{k=1}^N \rho_{ik}.
\end{equation}
DMD spectrum $\rho_i$ and Koopman exponents ($\lambda_i(t_k) = \log{\rho_{ik}}$) is related by
\begin{equation}
\log{\rho_i} = \frac{\lambda_i(t_1) + \lambda_i(t_2) + \cdots + \lambda_i(t_N)}{N}.
\end{equation}
Eqn.~(\ref{eqn:dmdvsKoopman}) used the knowledge that Koopman modes are state invariant.  

Generally, DMD numerical algorithm works for the following nonlinear cases.

\begin{enumerate}
\item The transient spectrum: If the observation interval is much smaller than the characteristic time $T$ of the dynamics system, the DMD algorithm may be useful to compute the transient spectrums and modes.

\item The infinite spectrum: If the observation duration is large enough such that $\tau\rightarrow \infty$ in Eqn.~(\ref{eqn:spectruminfty}) is approximately satisfied, $\lambda(\infty, t_0)$ defined in Eqn.~(\ref{eqn:spectruminfty}) will be approximated. If the system is further UES (or UAS), these infinite spectrums are time-independent and can be computed correctly by the DMD algorithm.

\item The limit cycle system: As stated earlier, periodic systems have constant Koopman spectrums which are independent on observation time and observation duration. DMD algorithm will correctly reveal their Koopman modes and spectrums.

\end{enumerate}

\section{Numerical examples for local Koopman decomposition}

In this section, the DMD algorithm is applied to two examples to obtain Koopman decomposition, where the hierarchy structure of the Koopman spectrums and the proliferation rule will be examined. The two examples are associated with the primary and secondary instability of wake past a cylinder.

\begin{figure}
\centering
\begin{overpic}[width=1.0\textwidth]{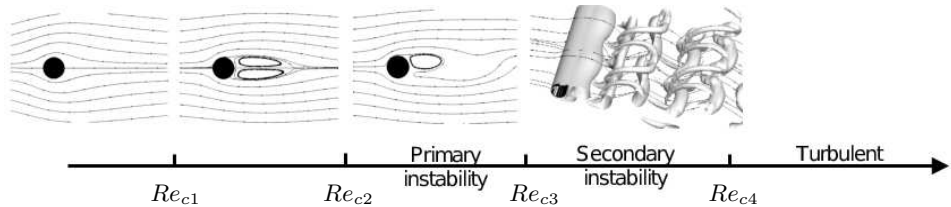}
\put(55,0){$Re_{c1}$}
\put(120,0){$Re_{c2}$}
\put(190,0){$Re_{c3}$}
\put(265,0){$Re_{c4}$}
\end{overpic}
\caption{Cylinder wake for different Reynolds number.} \label{fig:cylinderRe}
\end{figure}

The flow passing a fixed cylinder is chosen for its simple configuration, geometry and rich dynamics. The Reynolds number $Re = \frac{UD}{\nu}$ influences the instability of wake, where $U$ is the incoming flow velocity, $D$ is the diameter of the cylinder, and $\nu$ is the dynamic viscosity of the fluids. For low Reynolds number, such that $Re<Re_{c1}\approx 6$~\citep{jackson1987finite}, viscosity dominates. The flow is laminar, steady and does not separate from the cylinder. Above $Re_{c1}$, the flow separates from the cylinder surface and rejoins in the wake, creating a recirculating zone with two counter-rotating vortices. The flow is still symmetry, steady and laminar. Experiments show at around $Re_{c2}\approx 50$~\citep{tritton1959experiments,roshko1954development}, the wake breaks symmetry. The original two steady vortices shed alternatively off the cylinder, resulting the well-known K\'arm\'an vortex. Further increasing $Re$, it is observed at some point after $Re_{c3} \approx 190$~\citep{williamson1988existence} there is a sharp drop in the lift, drag as well as shedding frequency indicating a transition occured. At this Reynolds number, the original two-dimensional K\'arm\'an vortex begins to wobble in the spanwise direction, and eventually develops the three-dimensional waves. The above phenomena are summarized in figure~\ref{fig:cylinderRe}.

These bifurcation phenomena are attributed to the linear instability mechanism. In the range of $Re_{c2} < Re < Re_{c3}$, the linearized homogeneous Navier-Stokes equation around the steady base flow will provide a pair of unstable complex conjugate normal modes. When $Re$ exceeds the critical $Re_{c3}$, a three-dimensional secondary wave is formed and superimposed on the primary wave~\citep{landahl1972wave}. \citet{orszag1980transition} expanded the three-dimensional perturbation around the essential two-dimensional periodic base flow and obtained the Floquet system for the perturbation. Since the well-known instability mechanism and relatively simple flow dynamics, therefore, it is chosen as the study object.

Data for DMD analysis was obtained by numerically solving the governing equation. The incompressible Navier-Stokes equation was solved by the projection method~\citep{brown2001accurate}. A second-order central difference scheme was used for spatial discretization of the viscous term. Adams-Bashford scheme was employed for the convection terms and the Crank-Nicolson scheme was used for time advancement of viscous terms, and the sharp immersed boundary method (IBM)~\citep{mittal2008versatile} was implemented to handle the solids in the fluid. A non-conformal Cartesian mesh was used for the simulation. The staggered Cartesian grid was used to avoid the pressure-velocity decoupling. The numerical algorithm was the same as the one reported and tested in papers~\citep{yang2010numerical, xu2016embedded}.

\subsection{Koopman decomposition for primary instability}

\begin{figure}
\centering
\begin{subfigure}[b]{1.0\linewidth}
\includegraphics[width=0.9\linewidth]{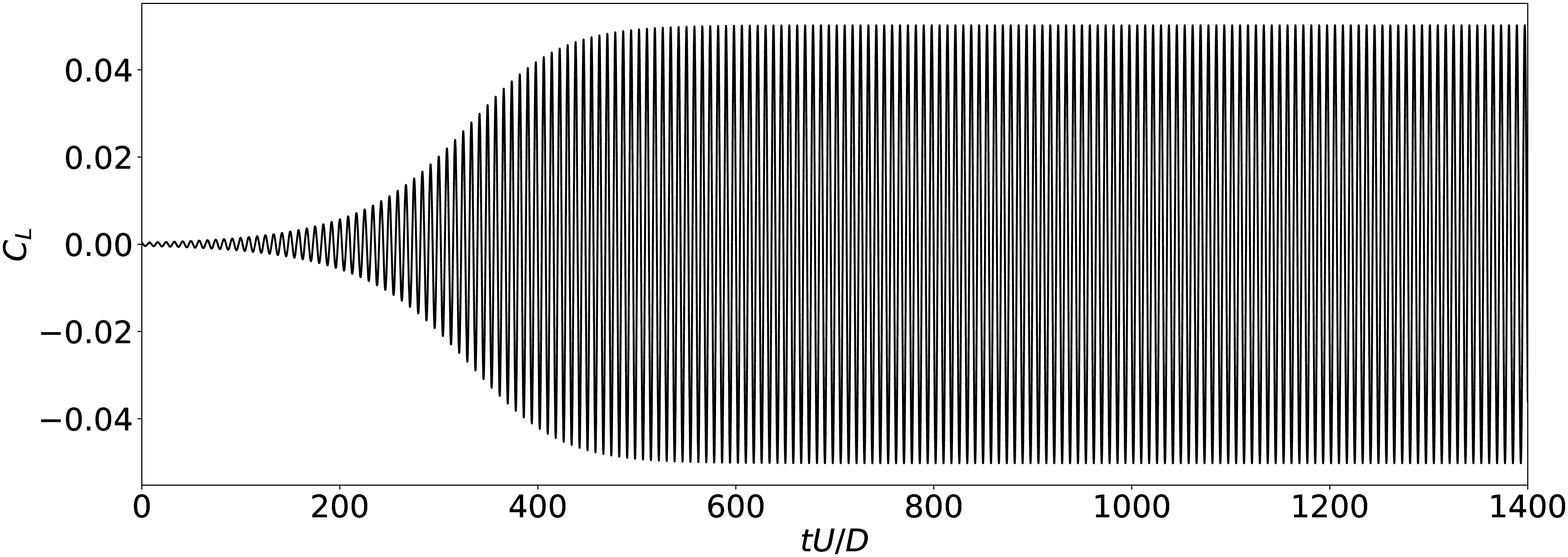}
\caption{The lift coefficient of 2D flow past fixed cylinder at $Re=50$.} \label{fig:2dcl}
\end{subfigure}
\begin{subfigure}[b]{0.49\linewidth}
\includegraphics[width=0.88\linewidth, trim={0 20 70 70}, clip]{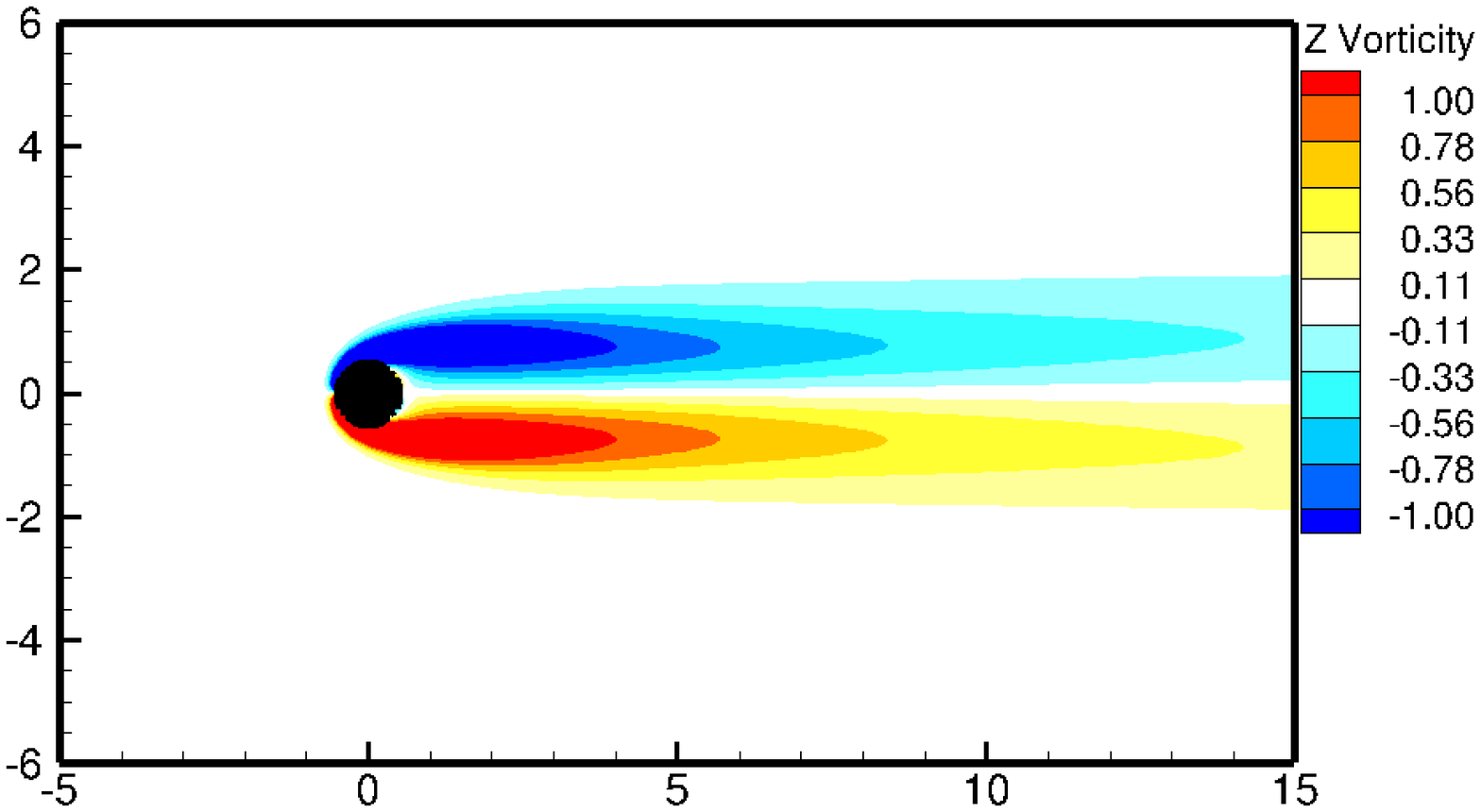}
\caption{initial field}
\end{subfigure}
\begin{subfigure}[b]{0.49\linewidth}
\includegraphics[width=1.0\linewidth, trim={0 20 0 70}, clip]{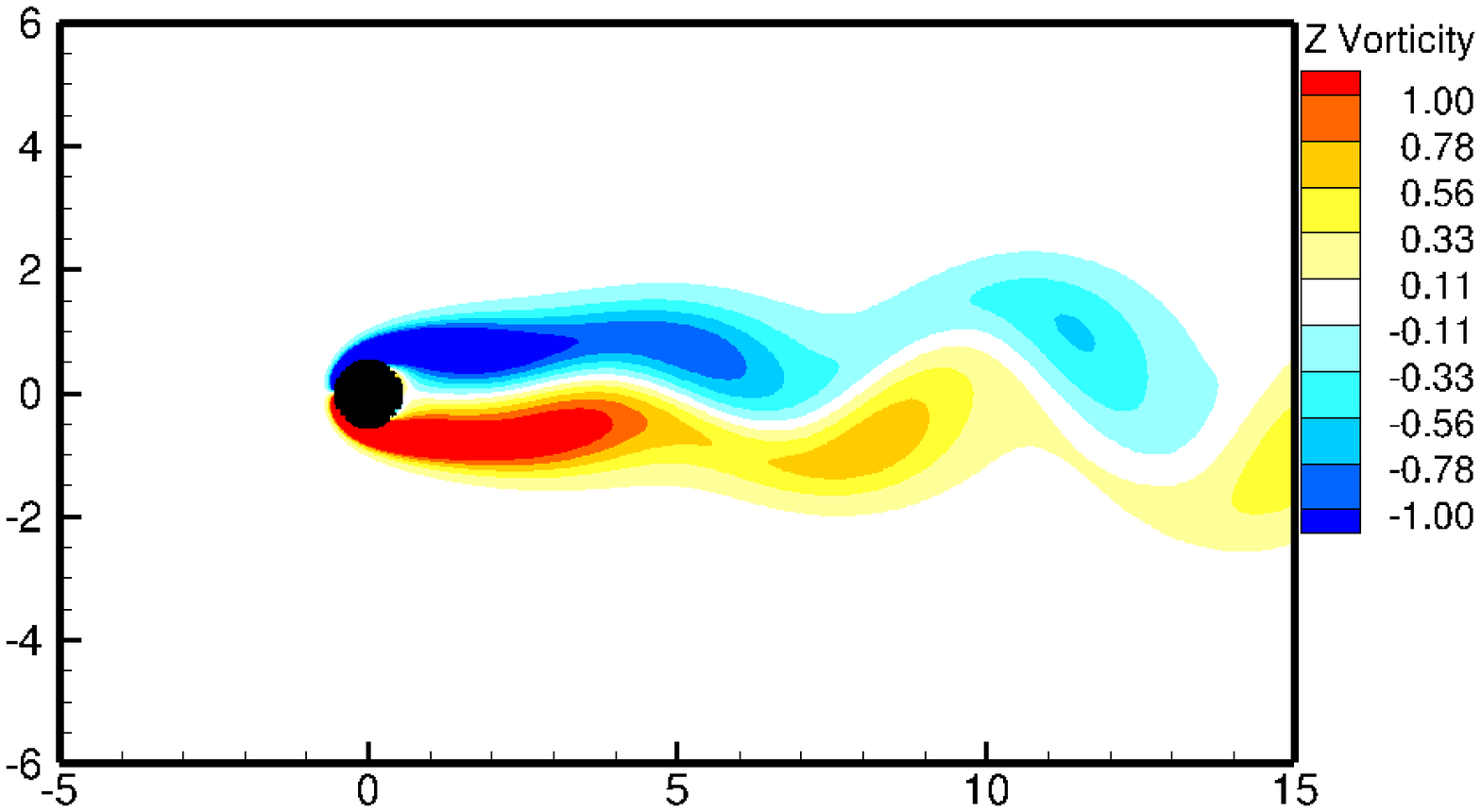}
\caption{fully-developed field}
\end{subfigure}
\caption{Numerical simulation of flow past fixed cylinder at $Re=50$. (a) The lift coefficient with respect to time. (b) (c) Vortex of initial and fully developed stage.}
\end{figure}

As stated earlier, at primary instability stage, the normal mode grows exponentially at a small magnitude. The growth of the normal mode saturates as its magnitude becomes big and the flow finally reaches periodic. Therefore, we divided the two-dimensional primary instability into two phases. The initial phase is for the initial growth of normal mode around the unstable equilibrium state. The final phase is for the saturation of the normal mode around the stable limit cycle solution. Koopman spectrums of each phase are studied separately. To help to illustrate the two phases the lift coefficient of the cylinder from the simulation is shown in figure~\ref{fig:2dcl}. The initial stage is chosen from  $tU/D=0\sim 167$, and the final stage is chosen from $tU/D=285\sim 1400$. 

\subsubsection{Koopman spectrums at the initial stage of primary instability}

\begin{figure}
\centering
\begin{subfigure}[b]{0.49\linewidth}
\includegraphics[width=1.0\textwidth]{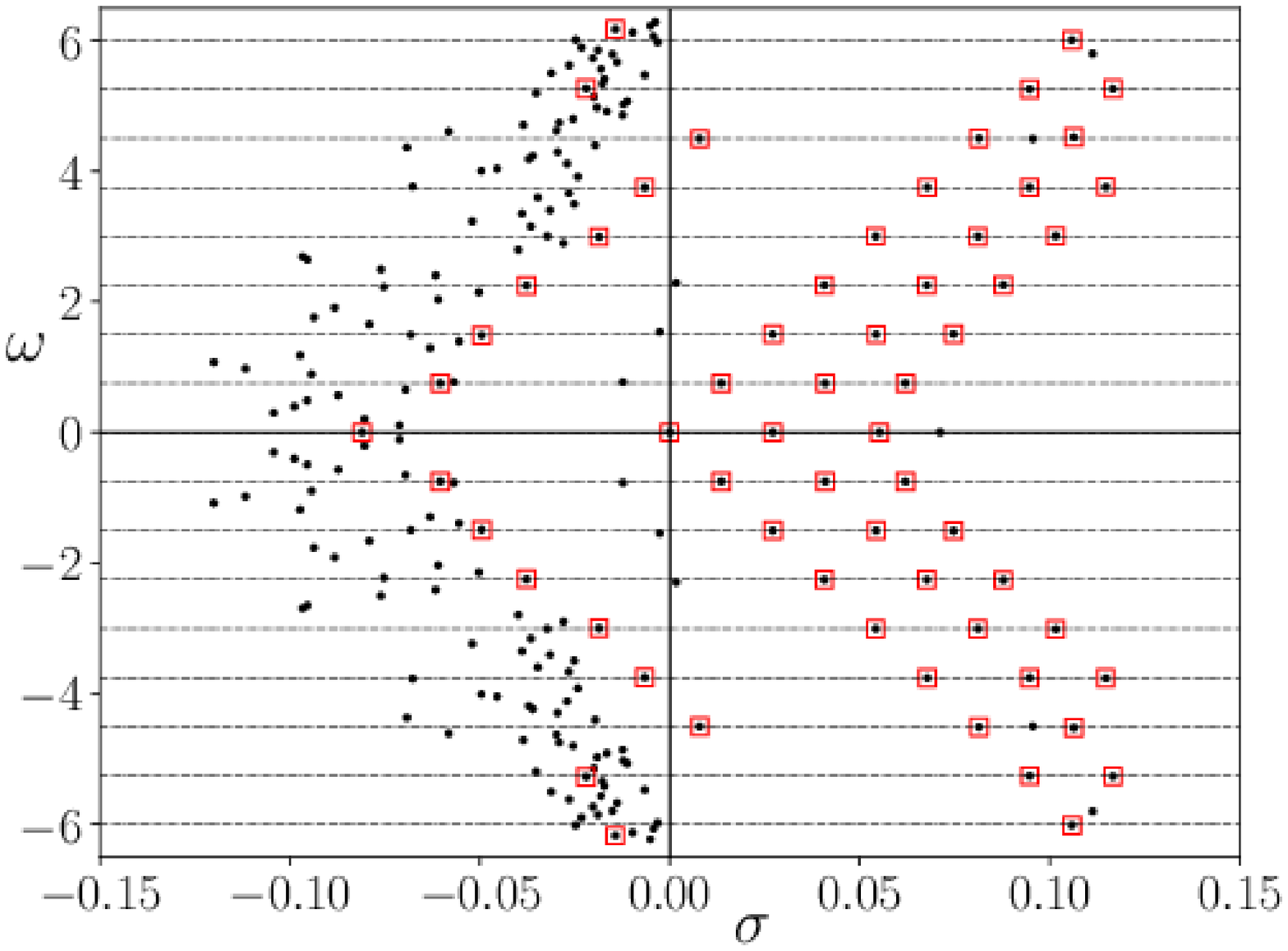}
\caption{initial stage \label{fig:spectruminit}}
\end{subfigure}
\begin{subfigure}[b]{0.49\linewidth}
\includegraphics[width=1.0\textwidth]{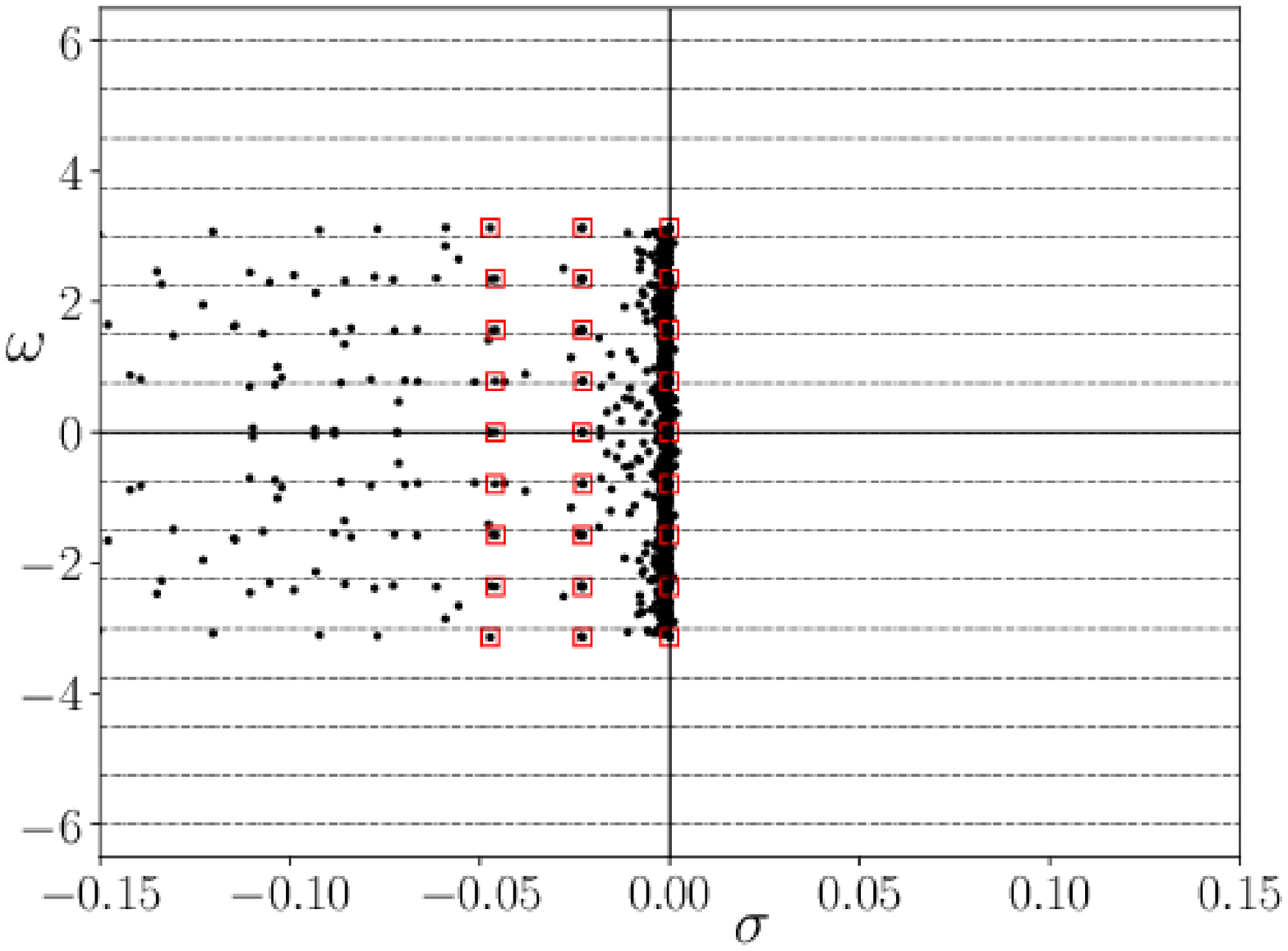}
\caption{finial stage \label{fig:spectrumfinal}}
\end{subfigure}
\caption{Koopman spectrums of primary instability of flow past fixed cylinder at $Re=50$. ($\bullet$) Black dots show all DMD modes. ({\color{red}{$\Box$}}) Red squares pick the modes with low residue.} \label{fig:primaryspectrum}
\end{figure}

The Koopman spectrums at the initial stage of primary instability are shown in figure~\ref{fig:spectruminit}.
The random perturbation at the incoming flow successfully excited the most unstable modes who have spectrums $(0.0136, \pm 0.751)$. These unstable normal modes and its derived high-order Koopman modes (whose spectrums form the triad-chain, all lie in the positive half plane) govern the wake. The triad-chain distribution is predicted by the proliferation rule. The spectrum $\lambda=0$ at the origin captures the base flow, the unstable equilibrium state. Besides these dominant modes, there is a stable mode $\lambda=-0.0810$ and two chains of modes starting from it, the latter are the predicted cross interaction by mode $\lambda=-0.0810$ and the unstable Koopman modes. Since the time interval between two consecutive snapshots is $\Delta t = 0.5$, the frequency is in the range $\frac{\text{Im}(\ln(\lambda))}{\Delta t}\in (-2\pi, 2\pi]$, thus frequency are truncated between $(-2\pi, 2\pi]$ as shown in figure~\ref{fig:spectruminit}.

\subsubsection{Koopman spectrums at the final stage of primary instability}

The Koopman spectrums for the final stage are shown in figure~\ref{fig:spectrumfinal}. Since the base flow is periodic, it is captured by the periodic modes ($\sigma=0$). Furthermore, it is known the nonlinear perturbation is dominated by a Floquet system. Therefore, Koopman modes are referred by Floquet exponents $\sigma$ but ignore their frequencies $\omega$. The least stable Floquet modes is $\sigma=-0.023$. $\sigma=-0.046$ captures the high order derived modes of $\sigma=-0.023$. The base spectrums, the spectrums for an underlined Floquet system, and the high-order derived spectrums in figure~\ref{fig:spectrumfinal} clearly show the hierarchy of Koopman spectrums for a periodic asymptotic system. Since the time interval between snapshots is $\Delta t = 1$ for this analysis, the frequencies are truncated into the range $\omega=\frac{\text{Im}(\ln(\lambda))}{\Delta t}\in (-\pi, \pi]$, as shown in figure~\ref{fig:spectrumfinal}.

Spectrums in Figure~\ref{fig:primaryspectrum} show clearly the hierarchy of Koopman spectrums, that is, the base spectrums and perturbation spectrums together form Koopman spectrums. And the triad-chain and lattice distribution of spectrum further confirm the proliferation rule.

\subsection{Koopman decomposition for secondary instability}

\begin{figure}
\centering
\begin{subfigure}[b]{0.9\linewidth}
\includegraphics[width=0.9\textwidth]{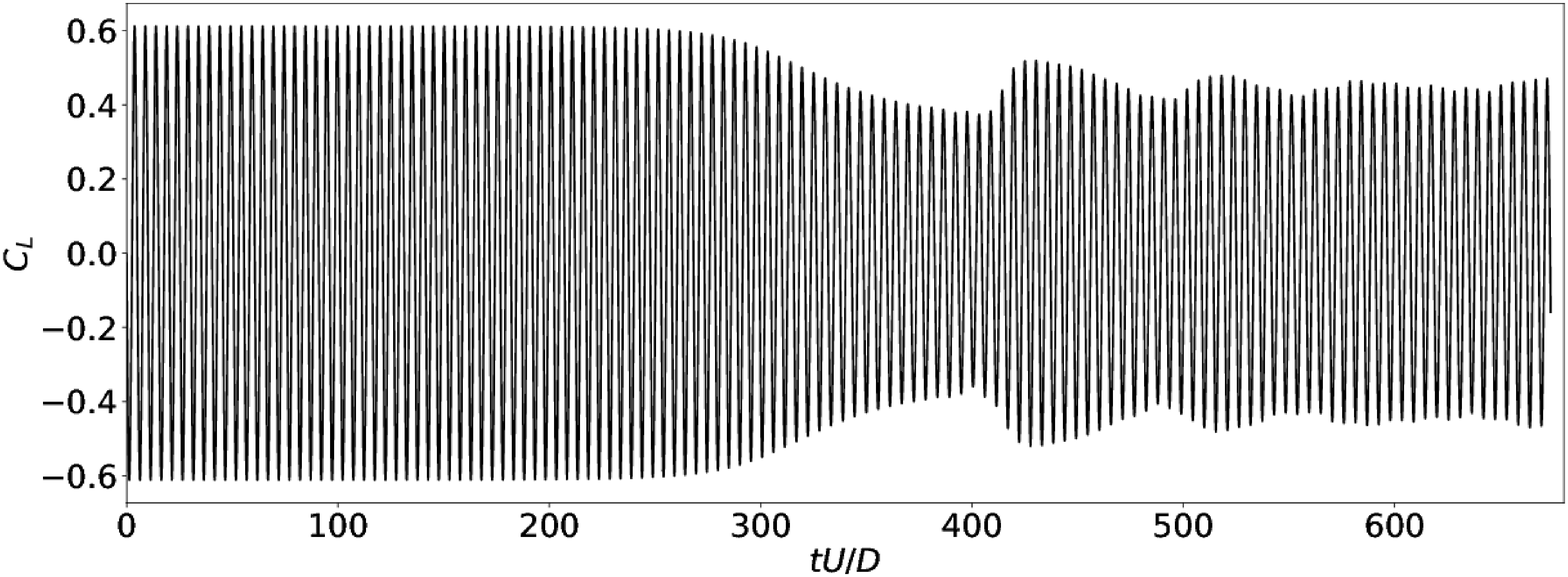}
\caption{History of $C_L$ over time.} \label{fig:3dcl}
\end{subfigure}
\begin{subfigure}[b]{1.0\linewidth}
\centering
\begin{overpic}[width=0.6\textwidth]{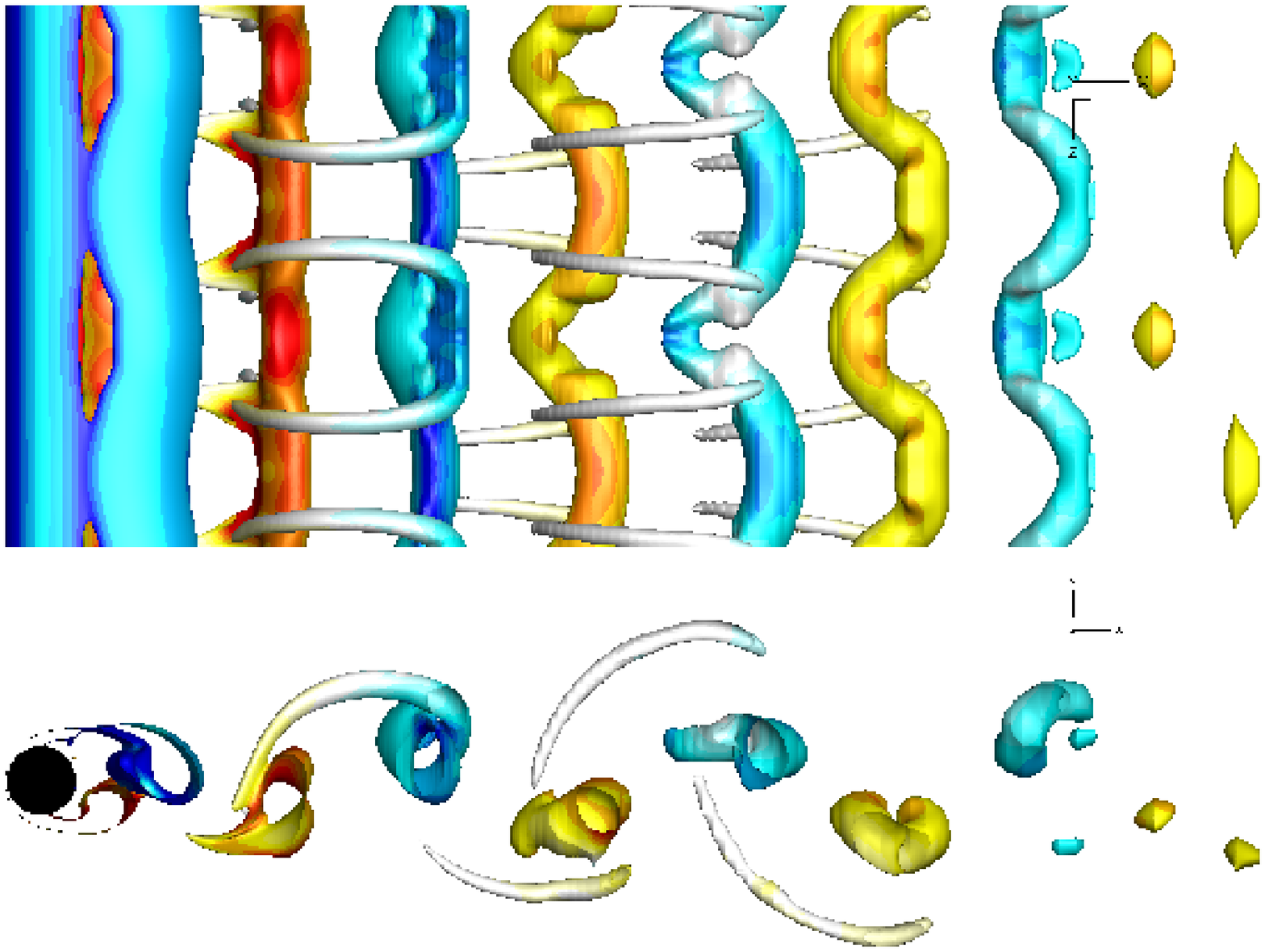}
\put(240,20){\includegraphics[width=0.1\textwidth, trim={30 0 490 37}, clip]{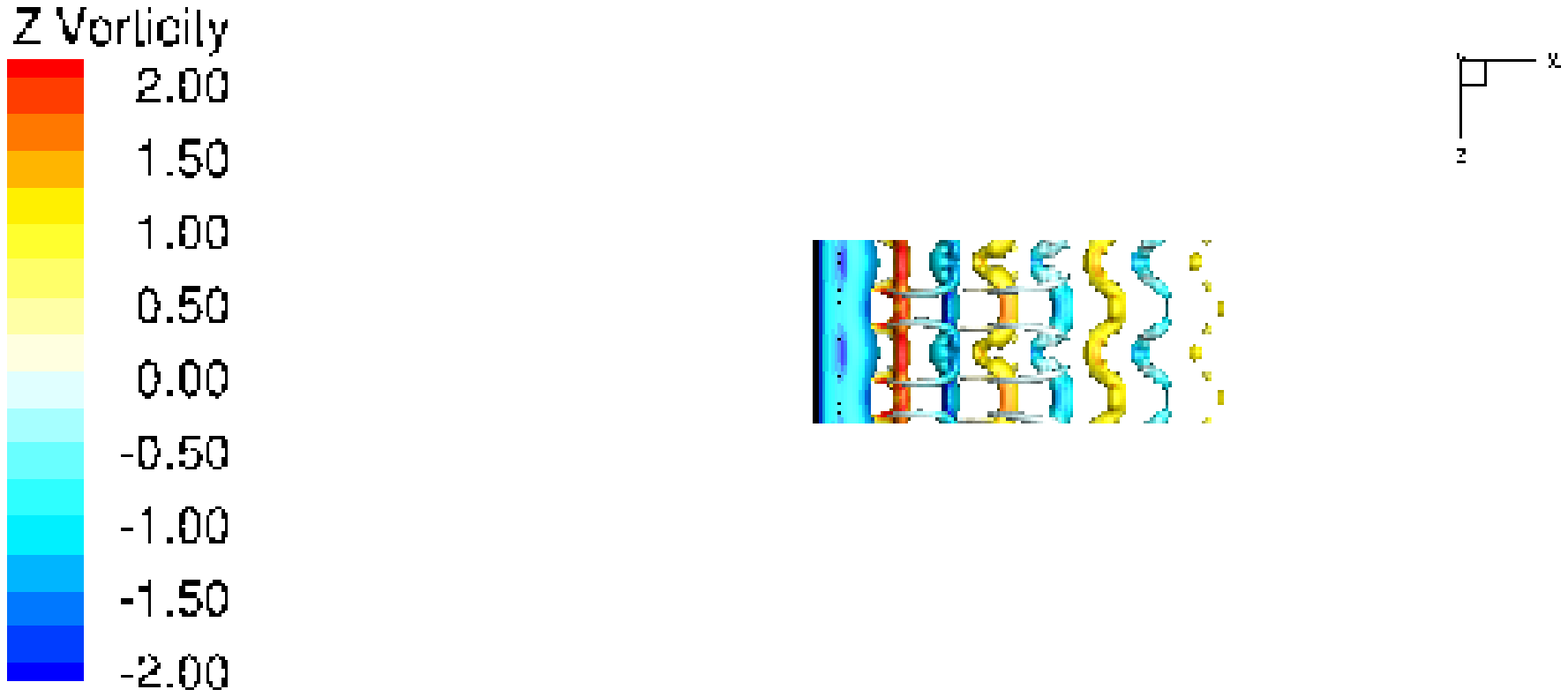}}
\put(255, 140){$\omega_z$}
\end{overpic}
\caption{Fully developed vortex after cylinder, top view and side view}
\end{subfigure}
\begin{subfigure}[b]{0.9\linewidth}
\includegraphics[width=0.49\linewidth]{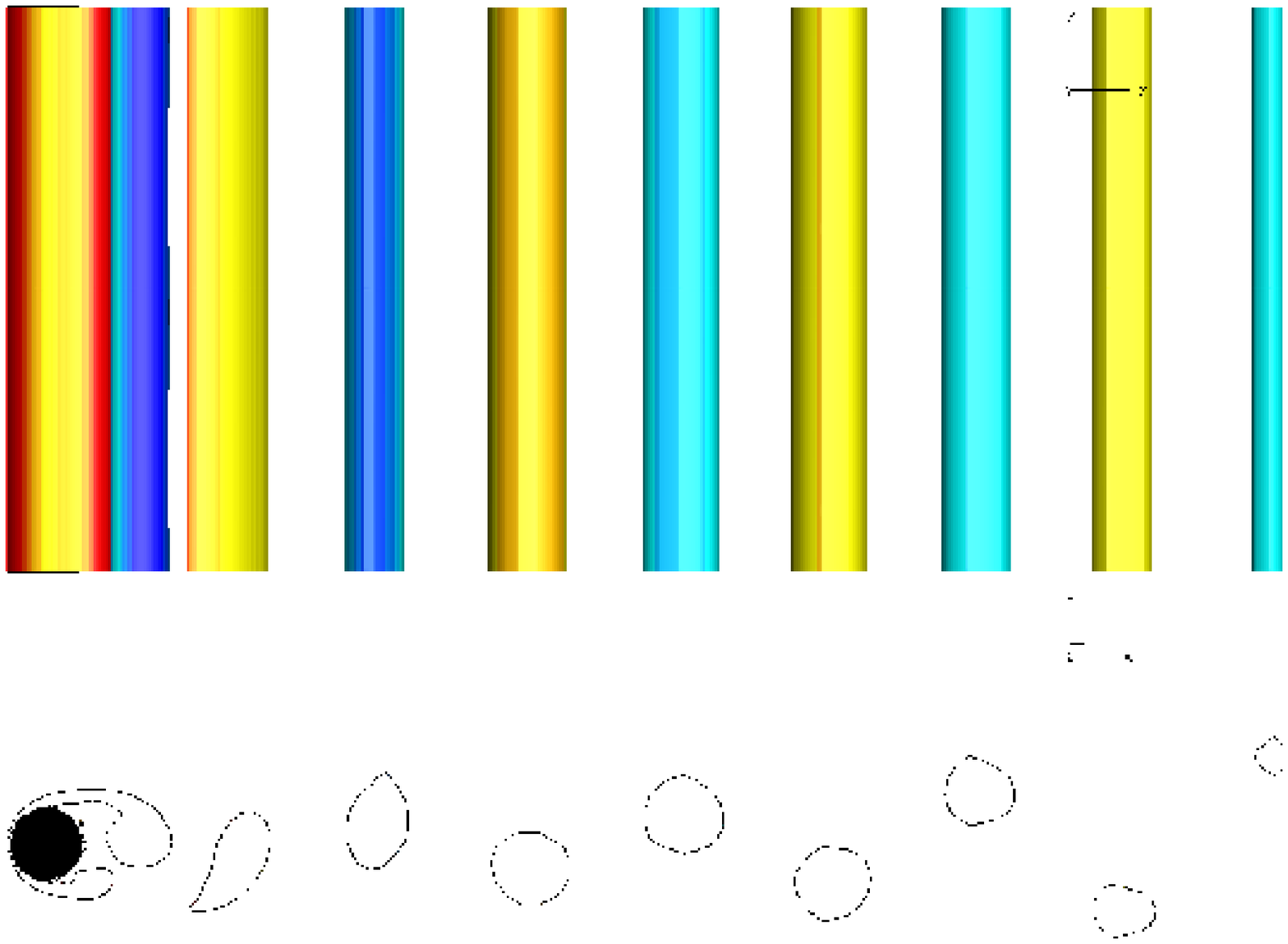}
\includegraphics[width=0.49\linewidth]{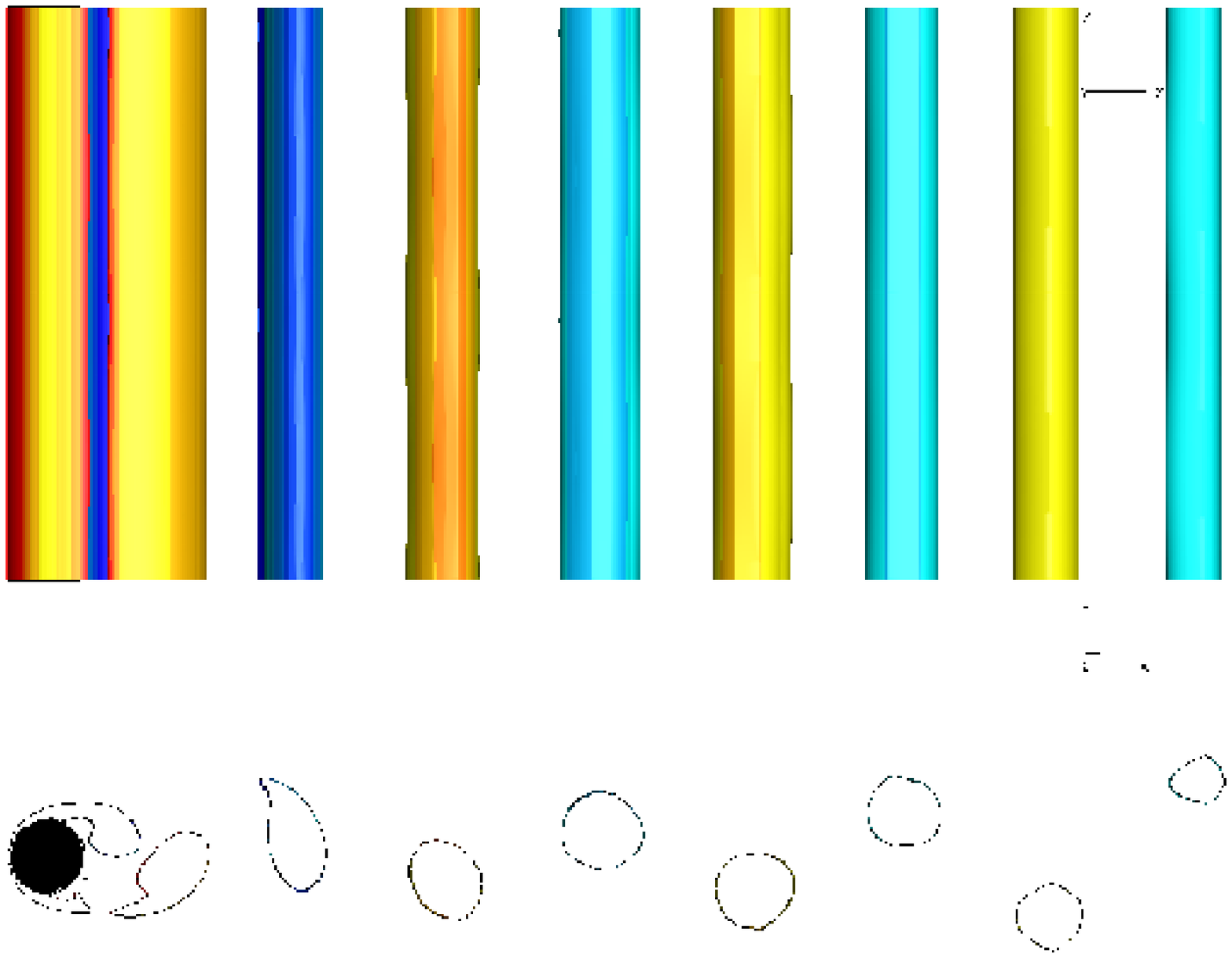}
%
\caption{The initial and final snapshots used for DMD analysis.}
\end{subfigure}
\caption{Numerical simulation of flow past fixed cylinder at $Re=200$. (a) The history of lift coefficient. (b) Fully developed vortex after cylinder at $Re=200$ with 3D simulation. Isosurface of Q-criterion of fully developed secondary instability wave of three-dimensional flow past fixed cylinder at $Re=200$, repeated 1 time (by periodicity) in spanwise direction to show the repeated structure. (c) The initial and final snapshots. The Q-criterion isosurface is colored by spanwise vorticity. A top view (x-z plane) and a side view (x-y plane) are shown.}
\end{figure}

To study the secondary instability, DMD algorithm is applied to the initial stage of secondary instability. Figure~\ref{fig:3dcl} shows the history of the lift coefficient $C_L$ during the numerical simulation. Data from $tU/D=110 \sim 320$ was chosen for DMD analysis. This data set contained 701 snapshots and the time interval of them was 0.3.

\begin{figure}
\centering
\includegraphics[width=0.7\textwidth]{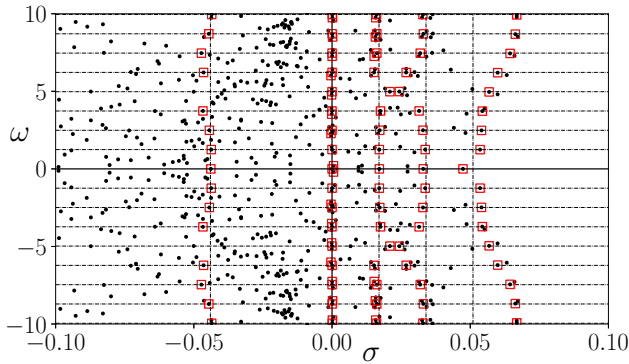}
\caption{The Koopman spectrum of three-dimensional secondary instability wave of flow past fixed cylinder at $Re=200$.} \label{fig:seondaryspectrum}
\end{figure}

Figure~\ref{fig:seondaryspectrum} shows Koopman spectrums for secondary instability. The lattice distribution of spectrums confirms the underlined system is a Floquet system. Black dots are the computed DMD eigenvalues, with red squares labeling low residue ones, see~\citep{zhang2019solving}. 

The spectrums with zero growth rate ($\sigma=0$) capture the periodic base flow. The unstable Floquet modes have a growth rate $\sigma=0.017$. The corresponding Floquet multiplier $e^{(\sigma T)} = e^{0.017/0.0181}=1.098$ is very close to $1.115\pm 0.005$ obtained by a direct Floquet analysis~\citep{abdessemed2009transient}. High order derived modes $\sigma=0.034$ and $\sigma=0.052$ are also captured. Besides these, a stable Floquet mode with decaying rate $\sigma=-0.044$ is captured. This may be due to the initial perturbation added to excite Floquet system. 

DMD algorithm may not fully recover the exact spectral pattern predicted by the proliferation rule (the lattice distribution here). It is because the Koopman spectrums are local; however, DMD working on a piece of data gives a time-averaged approximation. If nonlinearity is strong, the discrepancy can be significant as seen in figure~\ref{fig:seondaryspectrum}.

\begin{figure}
\centering
\begin{subfigure}[b]{0.325\linewidth}
\includegraphics[width=1.0\textwidth]{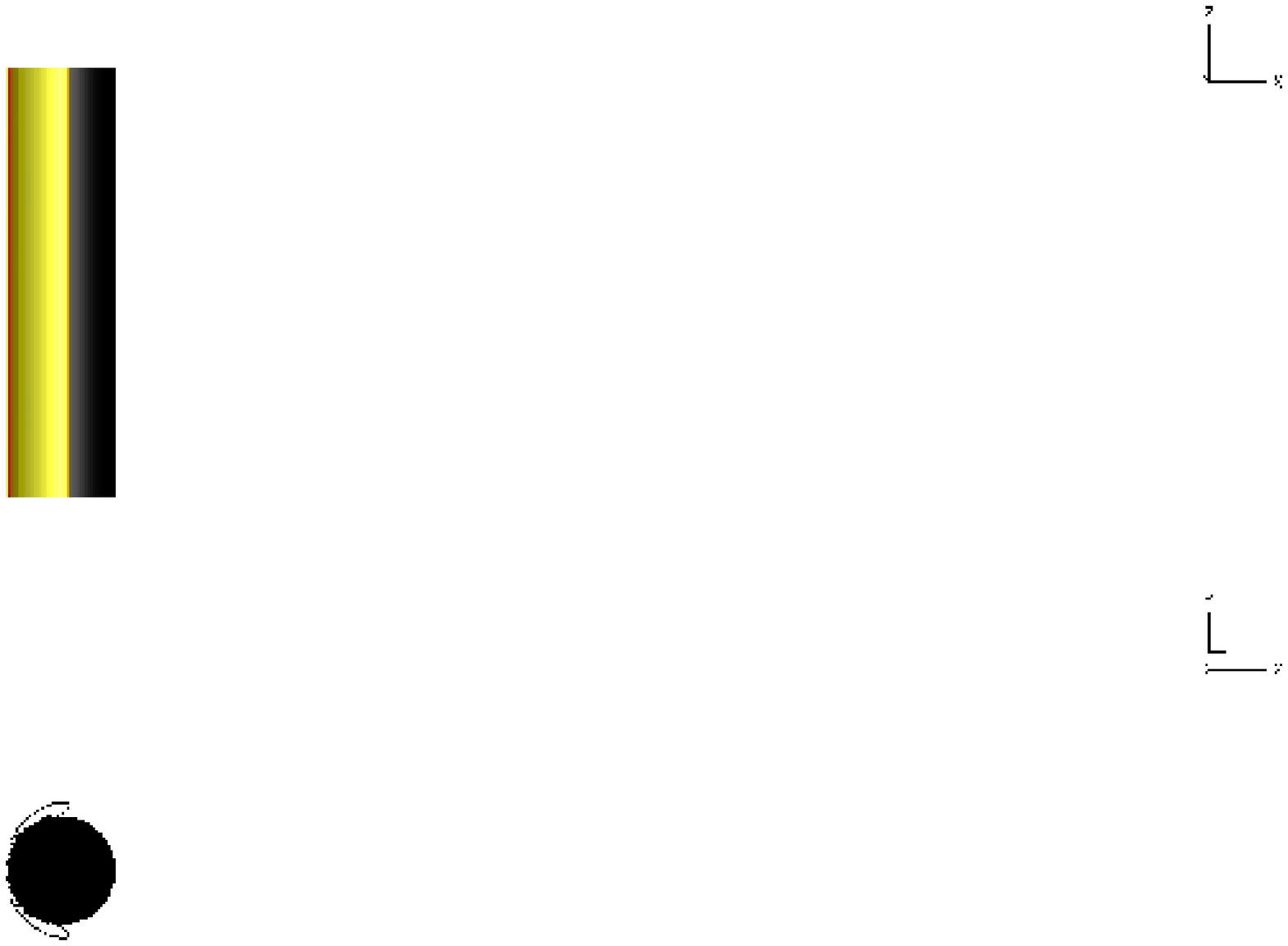}
\includegraphics[width=1.0\textwidth]{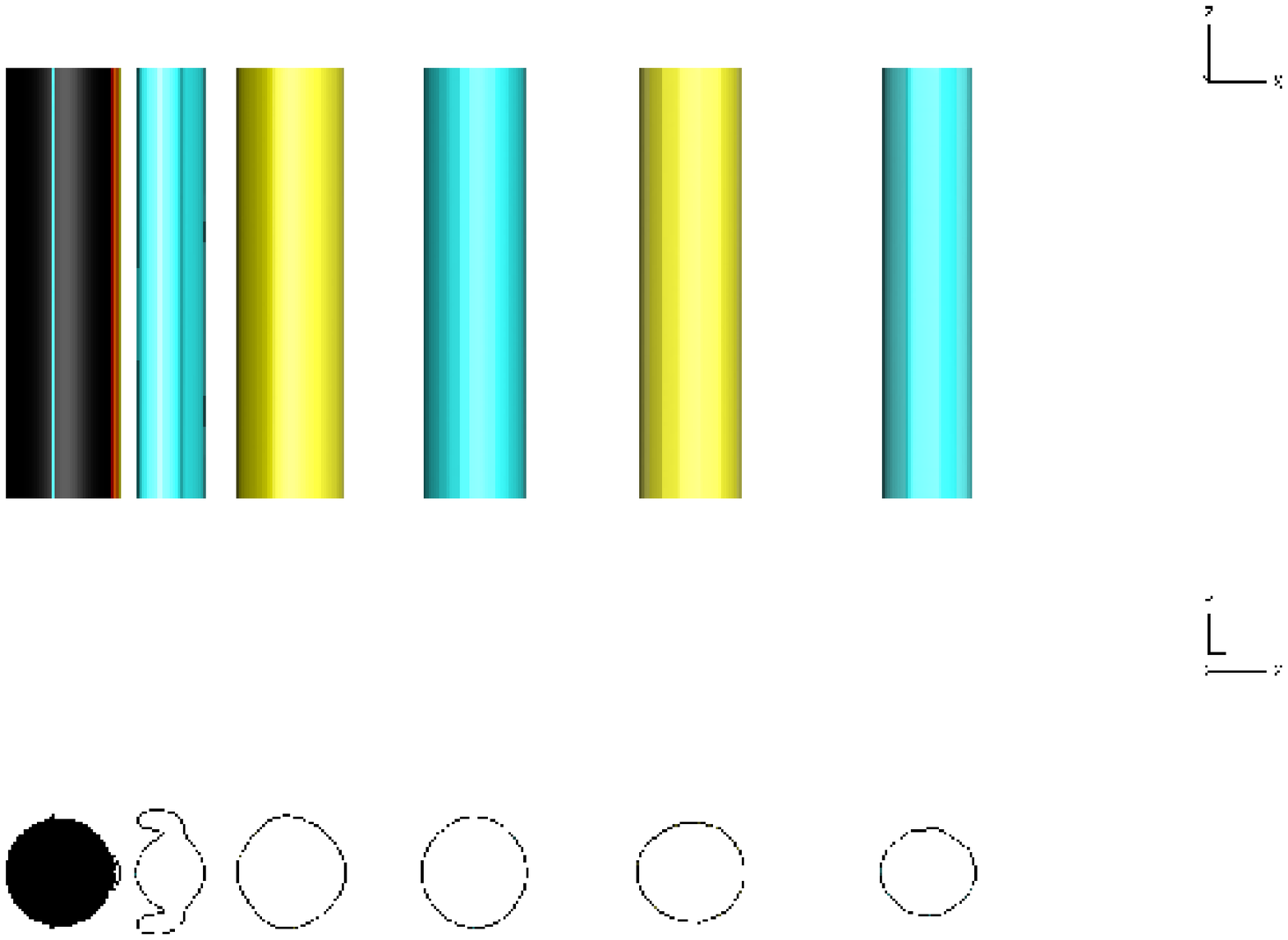}
\includegraphics[width=1.0\textwidth]{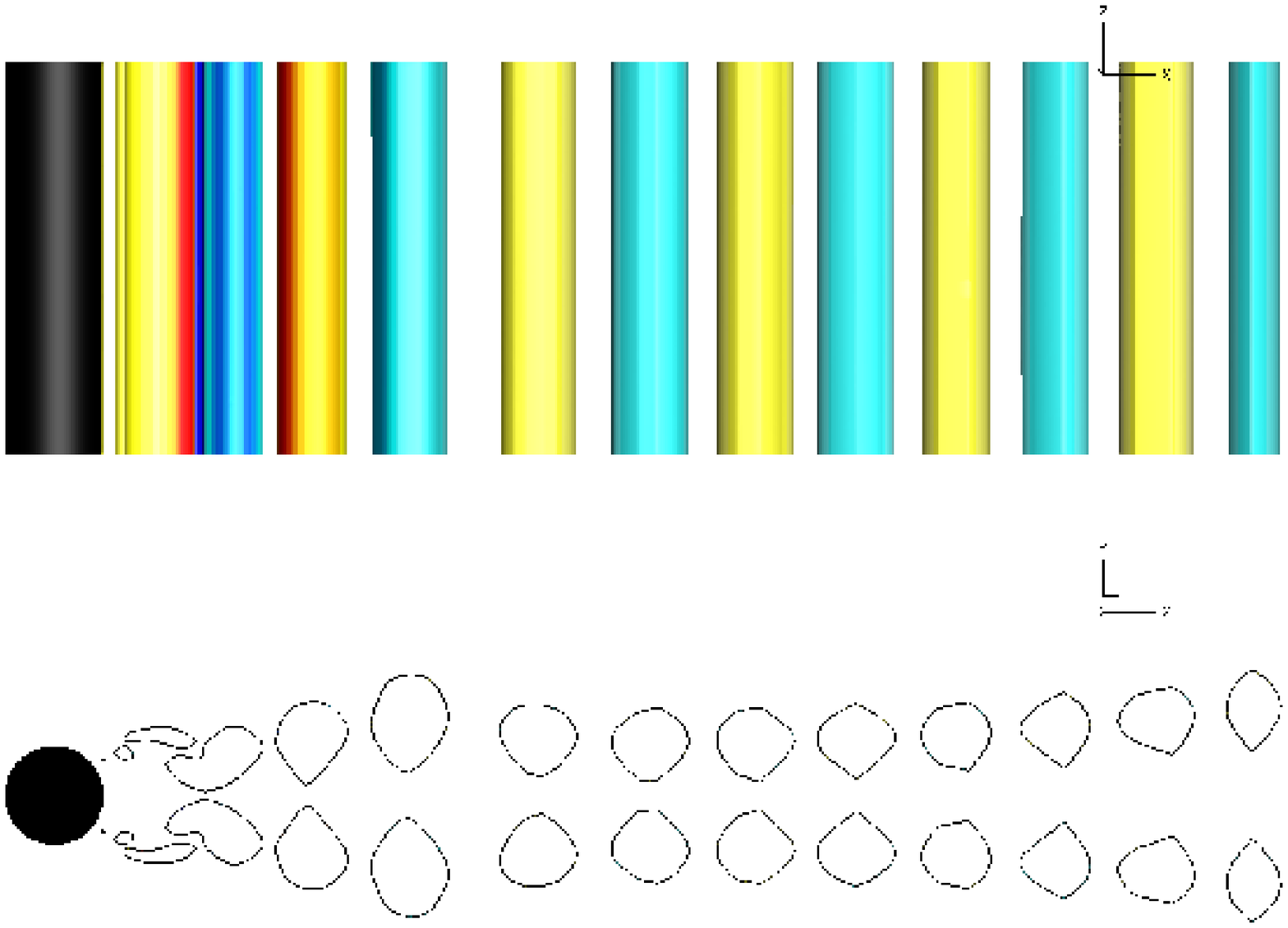}
\includegraphics[width=1.0\textwidth]{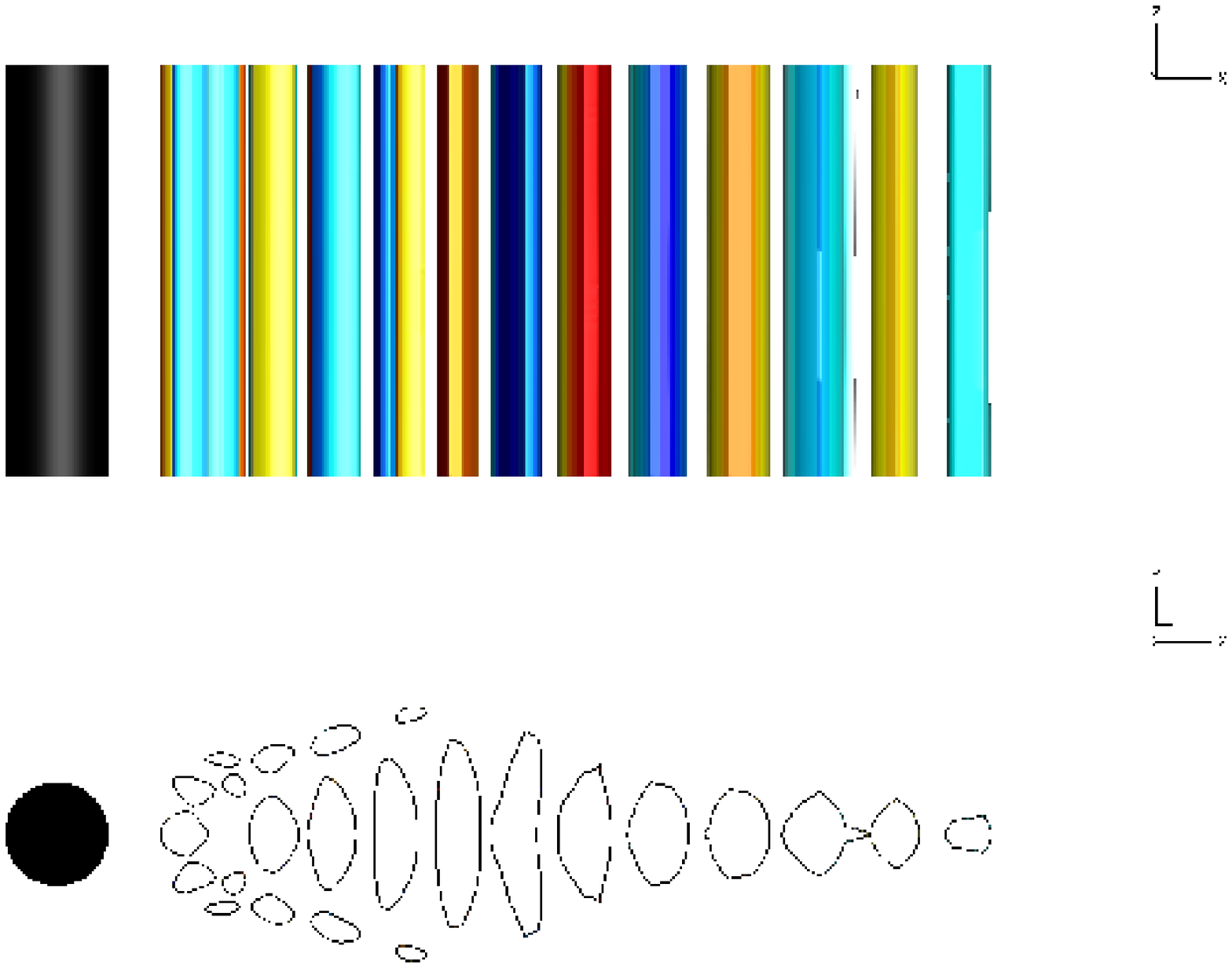}
\caption{$\sigma=0$}
\end{subfigure}
\begin{subfigure}[b]{0.325\linewidth}
\includegraphics[width=1.0\textwidth]{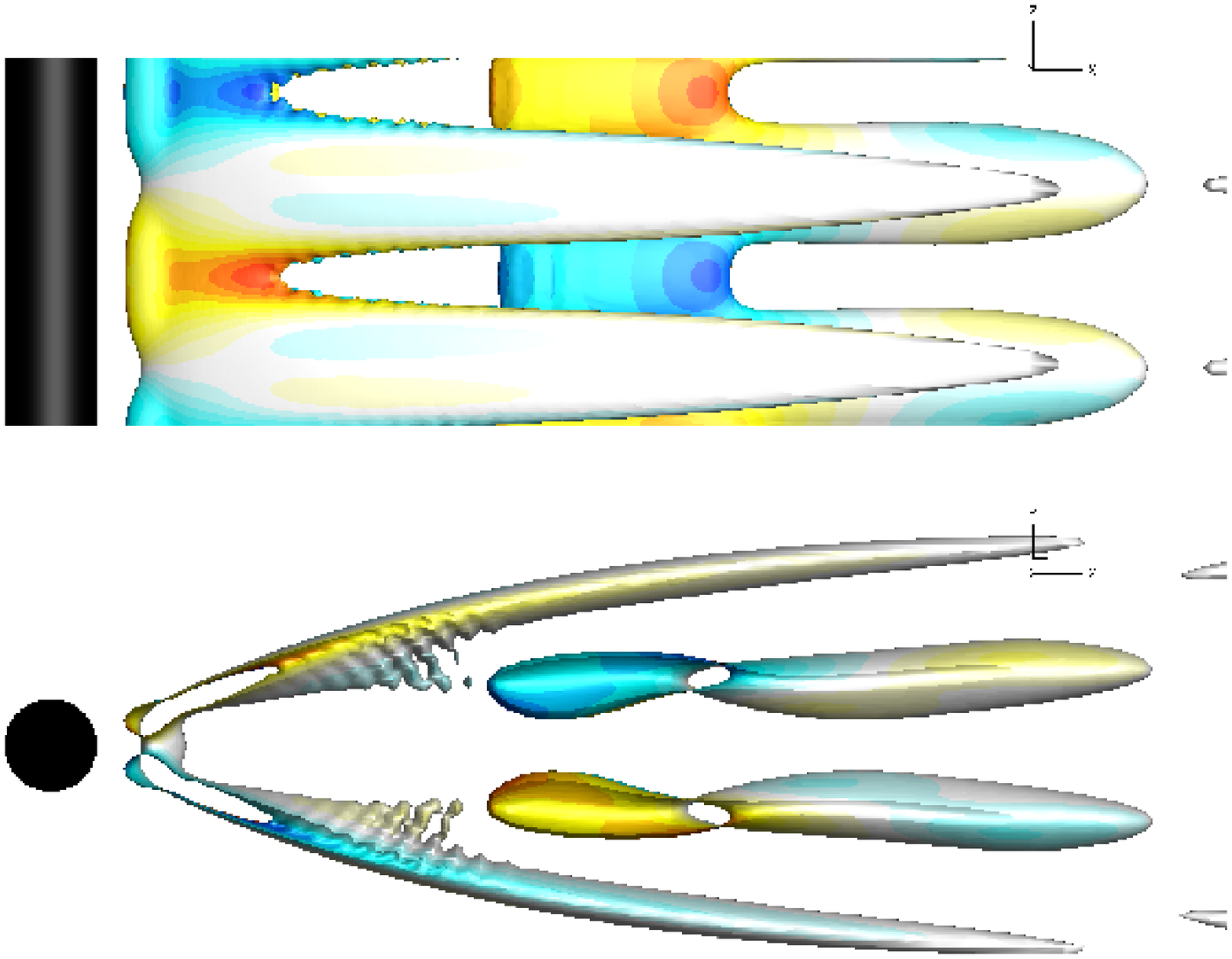}
\includegraphics[width=1.0\textwidth]{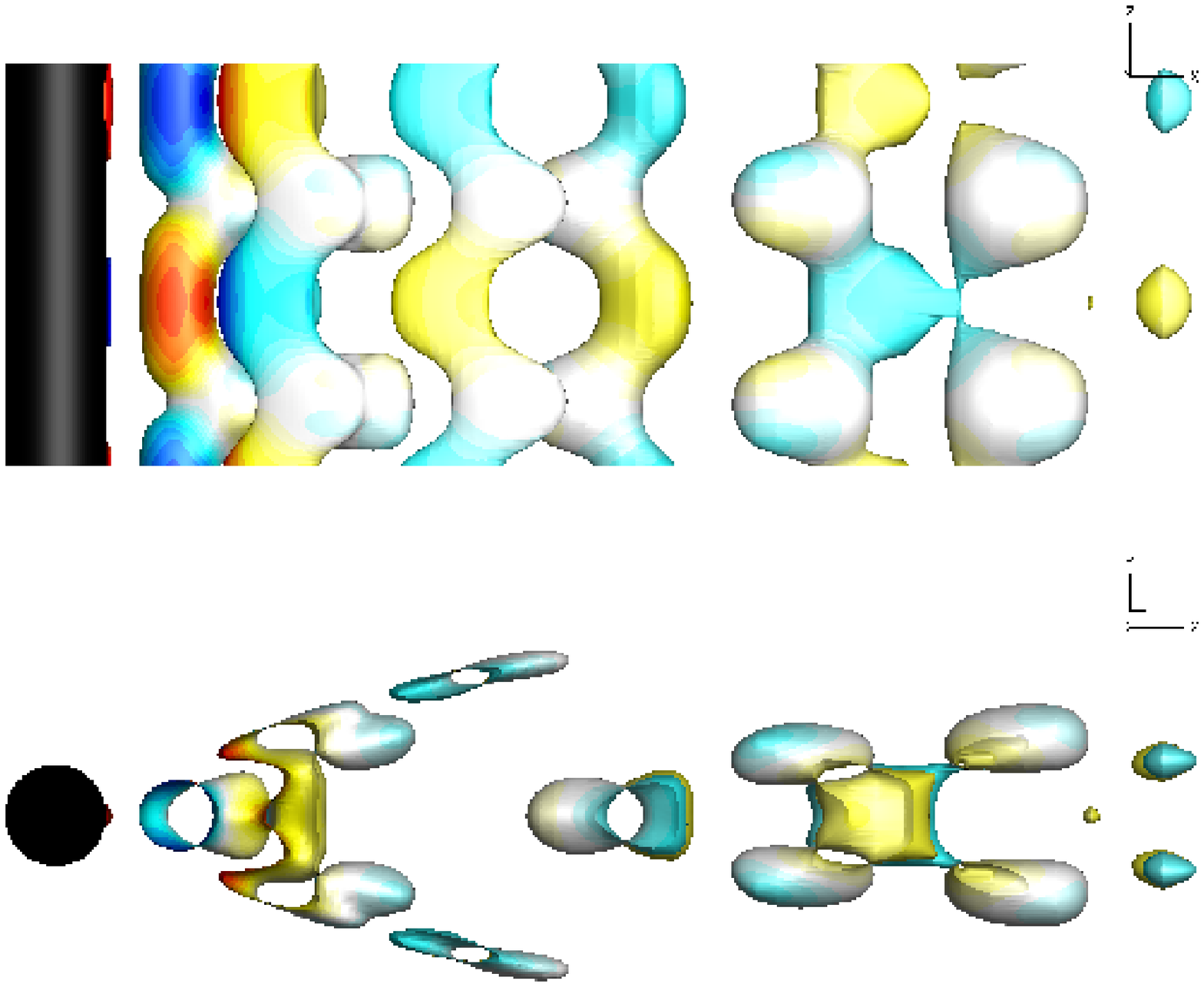}
\includegraphics[width=1.0\textwidth]{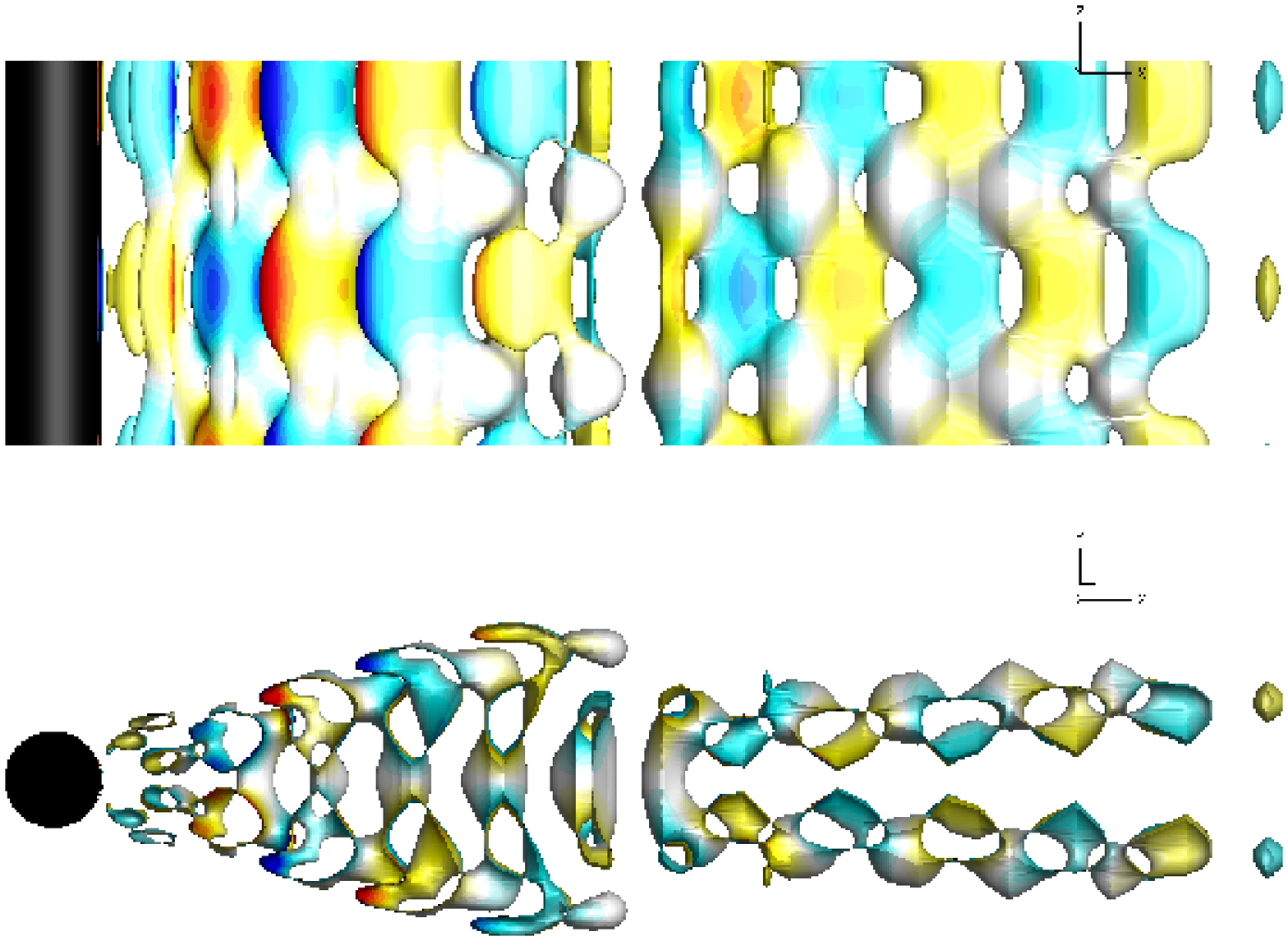}
\includegraphics[width=1.0\textwidth]{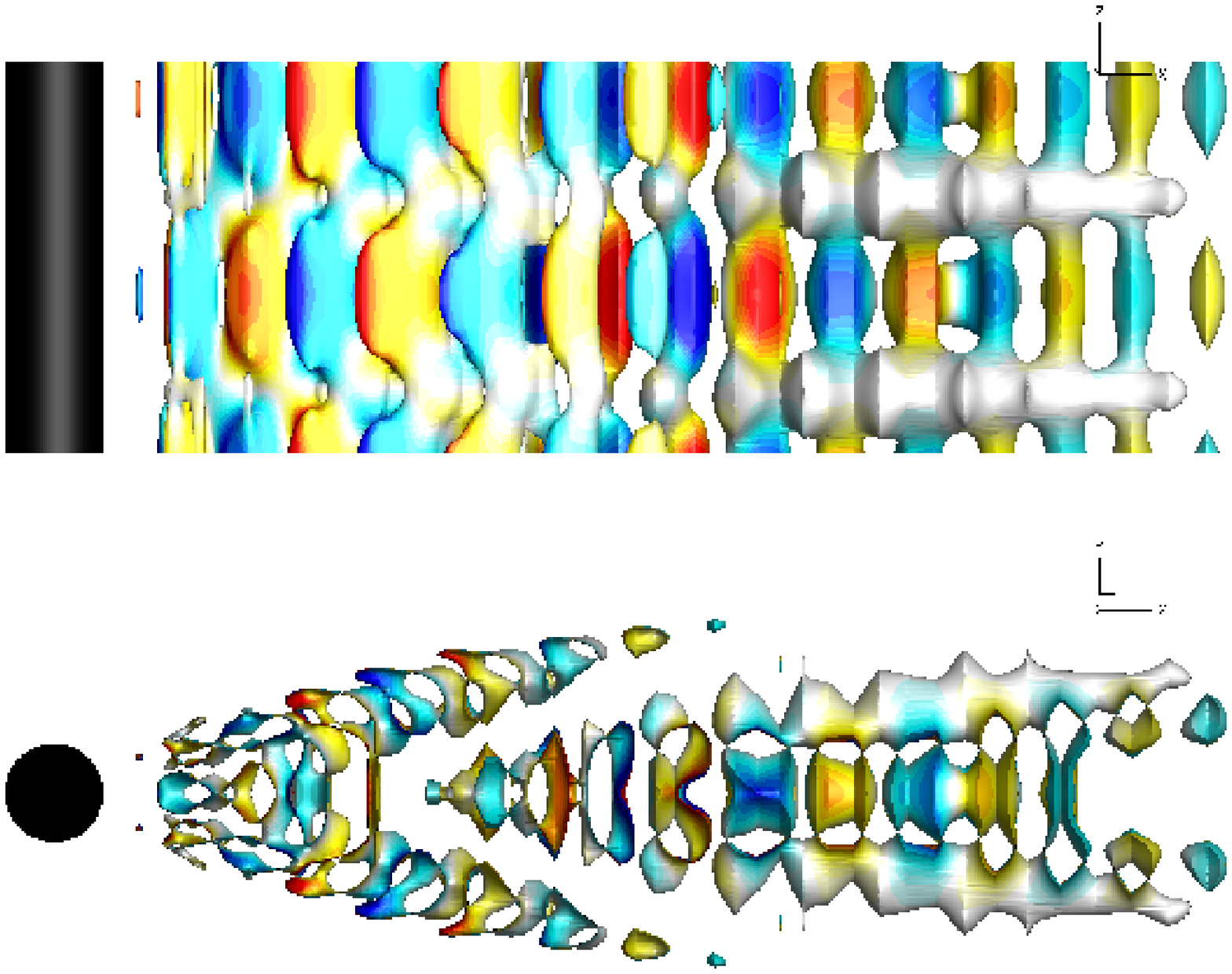}
\caption{$\sigma=0.017$}
\end{subfigure}
\begin{subfigure}[b]{0.325\linewidth}
\includegraphics[width=1.0\textwidth]{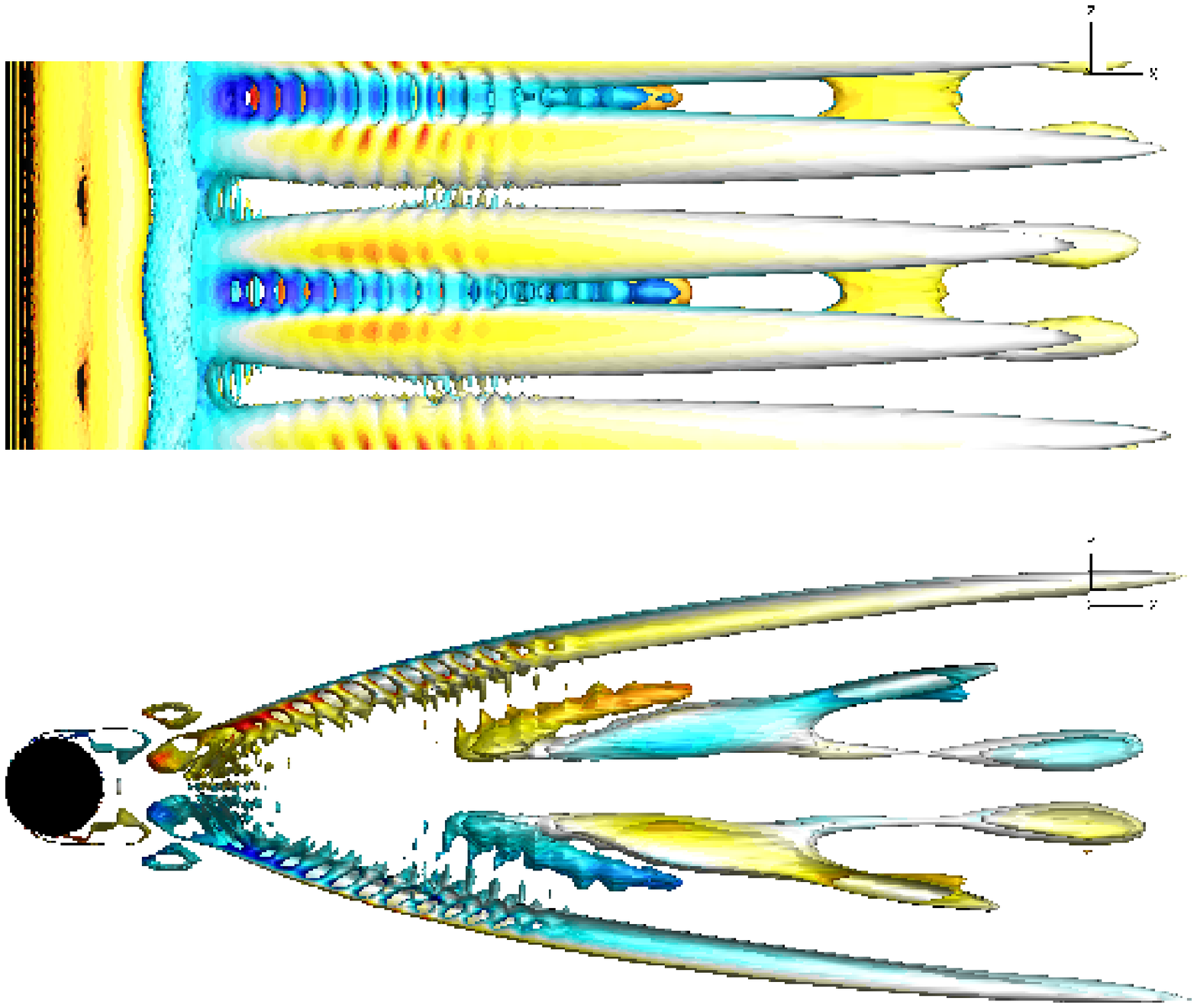}
\includegraphics[width=1.0\textwidth]{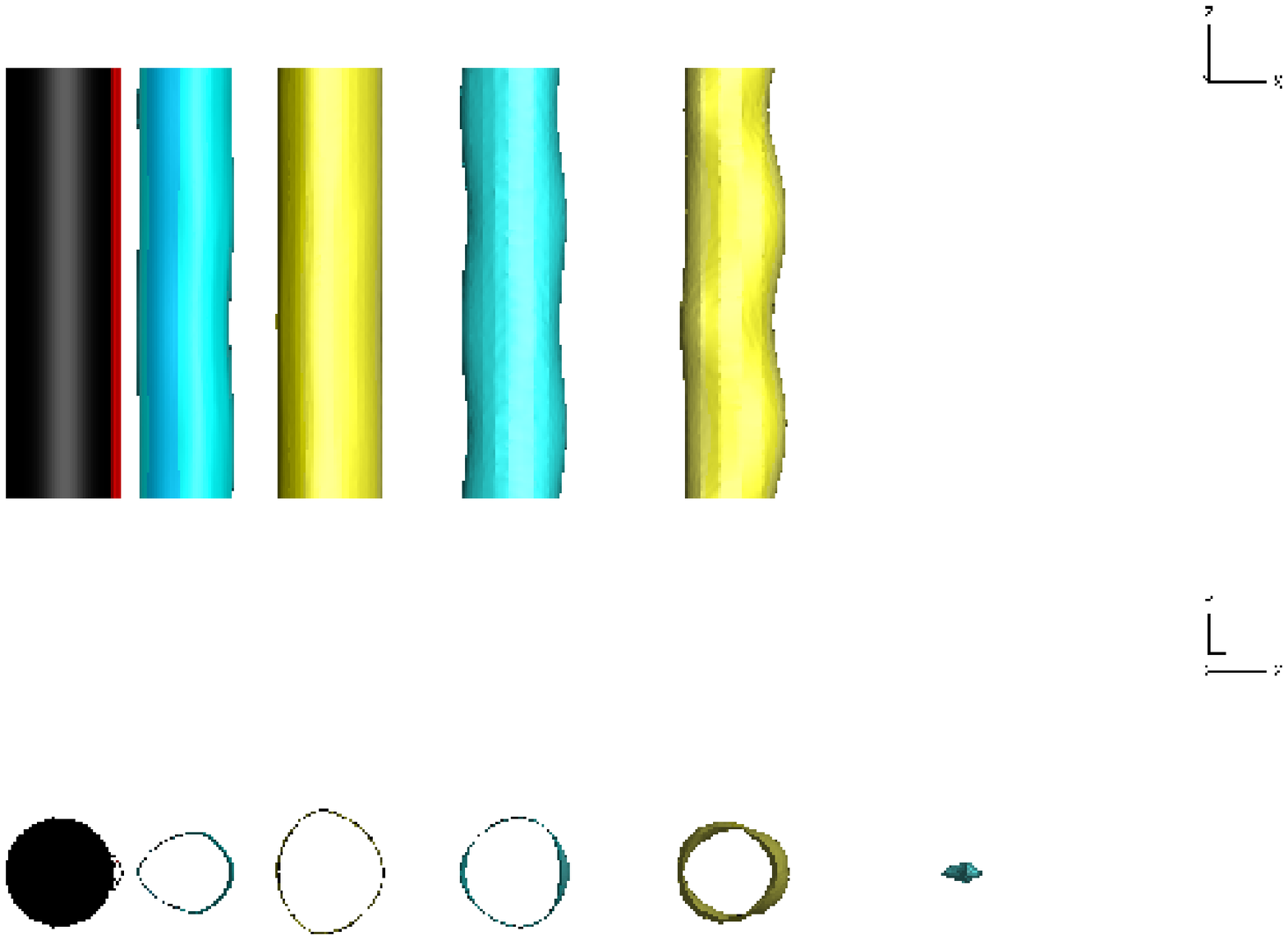}
\includegraphics[width=1.0\textwidth]{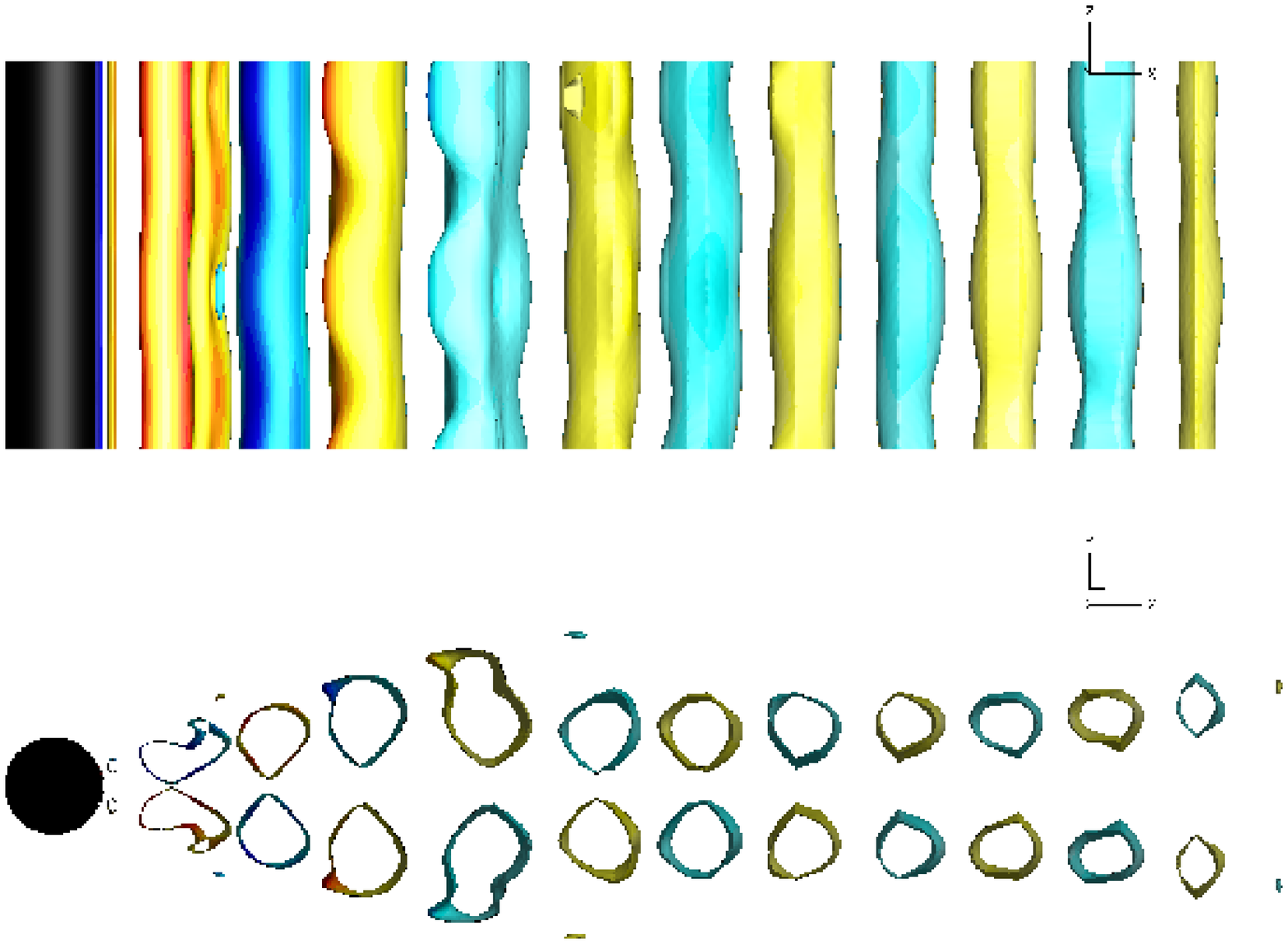}
\includegraphics[width=1.0\textwidth]{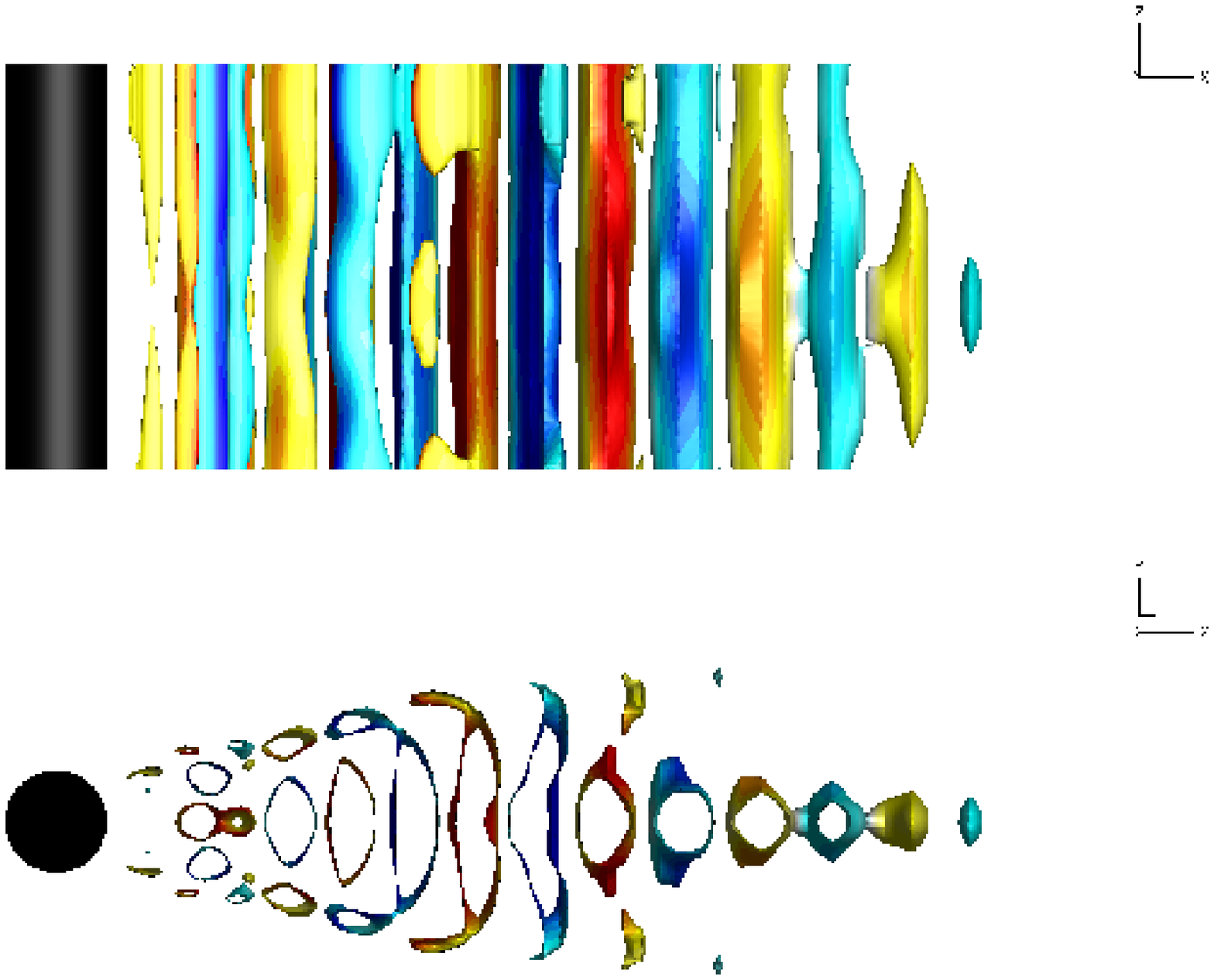}
\caption{$\sigma=0.034$}
\end{subfigure}
\caption{DMD modes for secondary instability of flow past fixed cylinder. (a) The first column shows the periodic base flow. (b) The second column shows Koopman modes corresponding to the most unstable Floquet modes (periodic part) decomposed into Fourier modes. (c) The third column shows the high order derived Koopman mode of the most unstable Floquet modes. The figures show the isosurface of Q-criterion and the isosurface is contoured by spanwise vorticity. For each mode, a top view (x-y plane) and a side view (x-z plane) are plotted.}\label{fig:3dmodes}
\end{figure}

Koopman modes are shown in figure~\ref{fig:3dmodes}. The periodic modes shown in the first column capture the base flow. They are essential two-dimensional modes, no wave in the spanwise direction. The most unstable Floquet mode is shown in column two. They have the same growth rate $\sigma=0.017$. A remarkable feature of them is that they contain only one wave in z-direction. The third column shows the high order derived mode of the unstable Floquet mode. These modes have a doubled growth rate of $\sigma=0.034$ and contain two waves in the spanwise direction. 

\section{Conclusions}

This paper studied the local Koopman spectral problem for nonlinear dynamic systems. The local Koopman spectral problem for various linear systems is first studied and found to be compatible with the linear spectral theory. For an LTI system, its linear spectrums constitute part of the Koopman spectrums. For an LTV system, the spectrums of the fundamental matrix are a subset of local Koopman spectrums. While for a periodic LTV system, it contains the Floquet spectrums or their augmentations with Fourier frequencies. Correspondingly, eigenvectors of these linear systems are also Koopman modes. For a nonlinear dynamic system, the local spectral problem is defined on the time-parameterized semigroup Koopman operator acting on it.

Several properties of Koopman spectrums are revealed. The recursive proliferation of eigenspaces into infinite dimensionl was found because of the nonlinearity of the systems. The hierarchy structure of Koopman spectrums of a nonlinear system is revealed, that is, the eigenspace is decomposed into the base and perturbation. The continuity of local Koopman spectrums is studied and found to be conditional continuous for nonlinear systems. The continuity property extends the knowledge of local dynamics to the global manifold. As a result, the Koopman modes are found to be state independent. 

Primary and secondary instability of flow passing a fixed cylinder are numerically analyzed using the DMD algorithm. The triad-chain or the lattice distribution of spectrums confirmed the proliferation rule and hierarchy structure. The obtained Koopman modes capture the main flow structures.

\section*{Acknowledgements}

The authors gratefully appreciate the support from the Army Research Lab (ARL) through Micro Autonomous Systems and Technology (MAST) Collaborative Technology Alliance (CTA) under grant number W911NF-08-2-0004.

\section*{Declaration of interests}

The authors report no conflict of interest.

\appendix
\section{Koopman decomposition of Periodic LTV systems} 
\label{sec:periodicLTV}


For a periodic LTV system
\begin{equation} \label{eqn:periodic}
\dot{\boldsymbol{x}} = A(t) \boldsymbol{x}, \quad \text{where } A(t+T) = A(t), 
\end{equation}
with initial condition $\boldsymbol{x}(t_0) = \boldsymbol{x}_0$.
$T$ is the smallest positive value for the period of the system. It is a Floquet system and its solution is described by the following theory~\citep{coddington1955theory}.

\begin{theorem}
If $\Theta$ is a fundamental matrix for system~(\ref{eqn:periodic}), then so is $\Psi$, where
\begin{equation*}
 \Psi(t) = \Theta(t+T), \quad (-\infty < t < \infty ).
\end{equation*}
Corresponding to every such $\Theta$, there exists a periodic nonsingular matrix $P$ with period $T$, and a constant matrix $R$ ($e^{TR}$ is called the monodromy matrix) such that
\begin{equation} \label{eqn:floquetsolution}
\Theta(t) = P(t) e^{tR}.
\end{equation}
\end{theorem}

The spectrum of monodromy matrix $e^{TR}$ is called \emph{Floquet multiplier}, and the spectrum of matrix $R$ is called \emph{Floquet exponent}.

Noticing $e^{0R} = I$, the fundamental matrix $\Phi(t,t_0)$ is  
\begin{equation} \label{eqn:floquetfundamental}
\Phi(t, t_0) = \Theta(t) \Theta(t_0)^{-1} = P(t)e^{tR} P(t_0)^{-1}.
\end{equation}

\subsection{Koopman spectrums for T-discretized systems}

Let $\tau = t_n-t_{n-1} = T$. The discrete form of the periodic LTV can be derived by
\begin{equation} \label{eqn:pdLTV}
\begin{split}
\boldsymbol{x}_{n+1} &= \Phi(t_{n+1},t_0) \boldsymbol{x}_0 \\
 &= P\left(t_0 + (n+1)T\right) e^{\left(t_0+(n+1)T\right)R} P(t_0)^{-1} \boldsymbol{x}_0 \\
 &= P(t_0) e^{TR} e^{(t_0+nT)R}P(t_0)^{-1} \boldsymbol{x}_0 \\
 &= P(t_0) e^{TR} P(t_0)^{-1} \left(P(t_0+nT) e^{(t_0+nT)R}P(t_0)^{-1} \boldsymbol{x}_0 \right) \\
 &= \Phi(t_{n+1}, t_n) \boldsymbol{x}_n.
\end{split}
\end{equation}
Here $\Phi(t_{n+1},t_n)$ is
\begin{equation} \label{eqn:mat4p}
\Phi(t_{n+1}, t_n) = P(t_0) e^{TR} P(t_0)^{-1}.
\end{equation}

Different from a general LTV system, a periodic LTV discretized at period $T$ has constant evoluting matrix similar to a LTI system. The corresponding Koopman eigenfunction is 
\begin{equation} \label{eqn:ltvkoopmaneigenfunction}
\phi_i(\boldsymbol{x}) = \left( \boldsymbol{x}, \boldsymbol{w}_i\right).
\end{equation}
Here $\boldsymbol{w}_i$ is the left eigenvector of matrix~(\ref{eqn:mat4p}). And
\begin{equation}
U \phi_i(\boldsymbol{x}) = \left( \Phi(t_n, t_{n-1})\boldsymbol{x}, \boldsymbol{w}_i \right) = \left( \boldsymbol{x}, \Phi(t_n, t_{n-1})^*\boldsymbol{w}_i \right) = \left( \boldsymbol{x}, \bar{\rho}_i\boldsymbol{w}_i \right) = \rho \phi_i(\boldsymbol{x}).
\end{equation}
Hence Floquet multiplier $\rho$ (the eigenvalue of $e^{TR}$) is also the Koopman multiplier. Floquet exponent $\lambda = \frac{\ln \rho}{T}$ is then the Koopman exponent. Koopman spectrums of T-discretized periodic LTV system are constant.

Koopman modes are given by the column of matrix $Q(t_0)$
\begin{equation}
Q(t_0) = P(t_0) V.
\end{equation}

\subsection{Koopman spectrums for continuous periodic LTV systems}  \label{sec:periodicLTVprove}

The above discrete spectrum considers the dynamics at every $T$ instance (one period). The continuous system~(\ref{eqn:periodic}) contains richer dynamic information.

From fundamental matrix~(\ref{eqn:floquetfundamental}), the solution reads
\begin{equation}
\boldsymbol{x}(t) = \Phi(t,t_0)\boldsymbol{x}_0 = P(t) e^{(t-t_0)R}P(t_0)^{-1} \boldsymbol{x}_0.
\end{equation}
Consider a simple case when $R$ is diagonalizable ($V^{-1}RV = \Lambda$).
\begin{equation} \label{eqn:solutionexpfloquet}
\boldsymbol{x}(t) = \underbrace{P(t)V}_{Q(t)} \underbrace{ e^{(t-t_0)\Lambda}}_{\text{Diag}} \underbrace{V^{-1}P(t_0)^{-1} \boldsymbol{x}_0}_{\boldsymbol{c}_0} = \sum_{i=1}^{n} c_i e^{\mu_i (t-t_0)}\boldsymbol{q}_i(t).  
\end{equation}
$P(t)$ is periodic, so are $Q(t)$ and $\boldsymbol{q}_i(t)$ (the column vector of $Q(t)$). $c_i$ is component of $\boldsymbol{c}_0$. Therefore, the Floquet solution contains exponential parts $e^{\mu_i t}$ and periodic parts $\boldsymbol{q}_i(t)$. The periodic $\boldsymbol{q}_i(t)$ can be expanded by Fourier series
\begin{equation} \label{eqn:solutionexpfloquetfourier}
\boldsymbol{x}(t) = \sum_{i=1}^n c_i e^{\mu_i (t-t_0)} \sum_{l=-\infty}^{\infty} \boldsymbol{q}_{il}e^{jl\omega t} = \sum_{i=1}^{n} \sum_{l=-\infty}^{\infty} c_i e^{(\mu_i+jl\omega) t} \boldsymbol{q}_{il} .
\end{equation}
Here $j=\sqrt{-1}$ and $\omega=\frac{2\pi}{T}$. The $e^{\mu_i t_0}$ is absorbed into $c_i$.

Solution in form~(\ref{eqn:solutionexpfloquet}) is given by a expansion of Floquet modes $\boldsymbol{q}_i(t)$ with exponential growth part $e^{\mu_it}$. Solution~(\ref{eqn:solutionexpfloquetfourier}) further expands the periodic Floquet modes $\boldsymbol{q}_i(t)$ by Fourier expansion. With this decomposition $\boldsymbol{q}_{il}$ is the new mode and $e^{(\mu_i+jlw) t}$ is the corresponding temporal growth part. The following section shows the two solution~(\ref{eqn:solutionexpfloquet},~\ref{eqn:solutionexpfloquetfourier}) define two sets of Koopman modes and eigenvalues.

\begin{figure}
\centering
\includegraphics[width=0.45\linewidth]{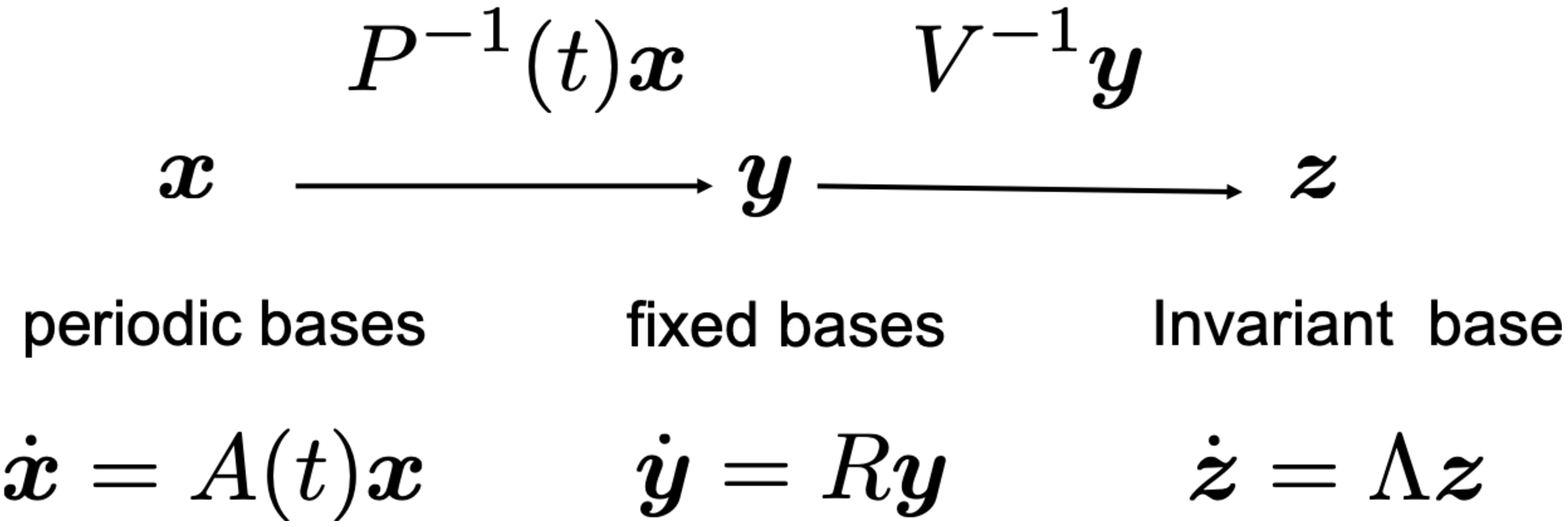}
\caption[Coordinate transformation of periodic LTV]{Coordinate transformation of periodic LTV.}\label{fig:coordtransform}
\end{figure}

Let
\begin{equation}
Q^{-1}(t) = \left[ \begin{array}{c} \boldsymbol{w}_1^H(t) \\ \vdots \\ \boldsymbol{w}_n^H(t)
\end{array}
 \right],
\end{equation}
and an observable 
\begin{equation} \label{eqn:eigenfunctionfloquet}
\phi_i(\boldsymbol{x}, t) = \left( \boldsymbol{x}, \boldsymbol{w}_i(t) \right) = \boldsymbol{w}_i^H(t) \boldsymbol{x}.
\end{equation}
Then
\begin{equation}
\begin{aligned}
U^{\tau} \phi_i(\boldsymbol{x},t) &= \phi_i(\boldsymbol{x}(t+\tau), t+\tau) = \boldsymbol{w}_i^H(t+\tau) \boldsymbol{x}(t+\tau) \\
 &= \boldsymbol{w}_i^H(t+\tau) P(t+\tau) V e^{\left((t+\tau)-t\right)R} V^{-1}P^{-1}(t) \boldsymbol{x}(t) \\
 &= \boldsymbol{w}_i^H(t+\tau) Q(t+\tau) e^{\tau R} \left( Q^{-1}(t) \boldsymbol{x}(t) \right) \\
 &= \boldsymbol{e}_i^H e^{\tau R} \boldsymbol{\phi}(\boldsymbol{x}, t) \\
 &= e^{\mu_i \tau}\phi_i(\boldsymbol{x}, t)
\end{aligned}
\end{equation}
proves $\phi_i(\boldsymbol{x}, t)$ is the Koopman eigenfunction with Koopman exponent $\mu_i$. In fact, the Koopman eigenfunctions defined by
\begin{equation}
\boldsymbol{\phi}(\boldsymbol{x}) = V^{-1}P^{-1}(t)\boldsymbol{x}
\end{equation}
is two consecutive coordinates transformations from $\boldsymbol{x}$ to a decoupled system, as illustrated by figure~\ref{fig:coordtransform}.

Further, an observable
\begin{equation} \label{eqn:eigenfunctionfloquetfourier}
\phi_{il}(\boldsymbol{x}, t) = e^{j l\omega t} \phi_i(\boldsymbol{x}, t)
\end{equation}
is also found to be the Koopman eigenfunction, since
\begin{equation}
\begin{aligned}
U^{\tau} \phi_{il}(\boldsymbol{x}, t) &= \phi_{il}\left(\boldsymbol{x}(t+\tau), t+\tau \right) = e^{j l\omega (t+\tau)}\phi_i \left(\boldsymbol{x}(t+\tau), t+\tau \right) \\
 &= e^{j l\omega (t+\tau)} \boldsymbol{w}_i^H(t+\tau) \boldsymbol{x}(t+\tau) \\
 &= e^{j l\omega (t+\tau)} \boldsymbol{w}_i^H(t+\tau) P(t+\tau) V e^{\left((t+\tau)-t \right)\Lambda} V^{-1}P^{-1}(t) \boldsymbol{x}(t) \\
 &= e^{j l\omega (t+\tau)} \boldsymbol{w}_i^H(t+\tau) Q(t+\tau) e^{\tau \Lambda} Q^{-1}(t) \boldsymbol{x}(t) \\
 &= e^{j l\omega (t+\tau)} \boldsymbol{e}_i^H e^{\tau \Lambda} \boldsymbol{\phi}(\boldsymbol{x}, t) \\
 &= e^{j l\omega (t+\tau)} e^{\mu_i \tau} \phi_i(\boldsymbol{x}, t) \\
 &= e^{(\mu_i+j l\omega)\tau} e^{jl\omega t} \phi_i(\boldsymbol{x}, t) \\
 &= e^{(\mu_i+j l\omega)\tau} \phi_{il}(\boldsymbol{x}, t) \\
\end{aligned}
\end{equation}
Therefore, $\phi_{il}(\boldsymbol{x}, t)$ and $\mu_i+j l\omega$ are the Koopman eigenfunction and exponent. Correspondingly,
the Koopman mode is obtained by expanding periodic matrix $Q(t)$ by Fourier expansions in the solution~(\ref{eqn:solutionexpfloquet})
\begin{equation}\label{eqn:invariantFloquetmode}
\begin{aligned}
U^{\tau}\boldsymbol{x}(t) &= \boldsymbol{x}(t+\tau) \\
 &= P(t+\tau) V e^{\left((t+\tau)-t \right)\Lambda} V^{-1} P^{-1}(t) \boldsymbol{x}(t) \\
 &= Q(t+\tau) e^{\tau \Lambda} Q^{-1}(t) \boldsymbol{x}(t) \\
 &= \left( \sum_{k=\infty}^{\infty} Q_k e^{j k\omega (t+\tau)} \right) e^{\tau \Lambda} \boldsymbol{\phi}(\boldsymbol{x}, t) \\
 &= \sum_{i=1}^n \sum_{k=-\infty}^{\infty} \boldsymbol{q}_{ik} e^{jk\omega(t+\tau)} e^{\tau \mu_i} \phi_i(\boldsymbol{x}, t) \\
 &= \sum_{i=1}^n \sum_{k=-\infty}^{\infty} \boldsymbol{q}_{ik} e^{(\mu_i+jk\omega)\tau } \left( e^{jk\omega t} \phi_i(\boldsymbol{x}, t)\right) \\
 &= \sum_{i=1}^n \sum_{k=-\infty}^{\infty} \boldsymbol{q}_{ik} e^{(\mu_i+jk\omega)\tau } \phi_{ik}(\boldsymbol{x}, t).
\end{aligned}
\end{equation}
By this expansion, the obtained Koopman modes $\boldsymbol{q}_{ik}$ is state-independent.

\bibliographystyle{jfm}
\bibliography{ms}

\end{document}